\documentstyle[floats,aps]{revtex}
\begin{document}

\title{Gravitational radiation from Schwarzschild black holes: \\the
second order perturbation formalism}

\author{Reinaldo J. Gleiser$^1$, Carlos O. Nicasio$^{1}$, Richard H.
  Price$^{2}$, Jorge Pullin$^{3}$} \address{1.  Facultad de
  Matem\'atica, Astronom\'{\i}a y F\'{\i}sica, Universidad Nacional de
  C\'ordoba,\\ Ciudad Universitaria, 5000 C\'ordoba, Argentina.}
\address{2. Department of Physics, University of Utah,
  Salt Lake City, Utah 84112.}  
\address{ 3. Center for Gravitational Physics and Geometry, Department
  of
  Physics,\\
  The Pennsylvania State University, 104 Davey Lab, University Park,
  PA 16802.} \date{May 6th, 1999 }

\maketitle
\begin{abstract}
The perturbation theory of black holes has been useful recently for
providing estimates of gravitational radiation from black hole
collisions.  Second order perturbation theory, relatively undeveloped
until recently, has proved to be important both for providing refined
estimates and for indicating the range of validity of perturbation
theory. Here we review the second order formalism for perturbations of
Schwarzschild spacetimes. The emphasis is on practical methods for
carrying out second order computations of outgoing radiation. General
issues are illustrated throughout with examples from ``close-limit''
results, perturbation calculations in which black holes start from
small separation.
\end{abstract}

\section{Introduction}
A common occurrence while studying a physical system is that the
situation that one is most interested in is too complicated to treat
in closed form.  Yet, for the very same system, it might be possible
to study in a closed form a situation that has a higher degree of
symmetry. An intermediate avenue has usually been to consider the
system in a regime that is more akin to the one of interest but that
departs ``slightly'' from the situation of high symmetry for which one
can solve exactly. One is therefore studying ``small perturbations''
of an exact solution. Considering the perturbations as small allows
them to be described by a linear theory. The resulting equations usually
simplify so much that they can be solved in situations without any
particular symmetry, provided they depart only slightly from an
exactly known configuration.

This approach is applicable in general relativity. The field equations of
the theory are a set of non-linear partial differential equations that are
not solvable in many situations of interest. In fact, state of the art
computers might not be able to evolve solutions far enough in time for the
physics of interest to be captured. Yet, if one assumes spherical
symmetry, an exact solution has been known since 1916. The exact solution
is the Schwarzschild solution that describes the geometry of a black hole.
As the concept of black hole came to better physical understanding in 
the 1950's, people started studying spacetimes that represented ``small
departures'' from a black hole geometry. Regge and Wheeler  \cite{rw} 
were the first to study these departures, and were able to find a clean
formulation for one-half of the degrees of freedom of the problem. The 
formulation for the other half had to wait until the work of Zerilli
 \cite{zprl}. Perturbations of flat spacetime had been considered earlier
on  \cite{LaLi}, and in fact Einstein himself discovered an approximate
version of the Schwarzschild solution as a perturbation of flat space. 

The motivation of Regge and Wheeler to study perturbations of black
holes was to assess the stability of the black hole solutions. Were
the black hole solutions ``an accident'' 
that arose as a consequence of the assumption of
spherical symmetry only, or did they persist if perturbed? The studies
of Regge and Wheeler and Zerilli, and later on of Price  \cite{Pr} and others
indeed showed that the black hole solutions are stable under small
perturbations. Stability was also the motivation of the early
second-order perturbations studies due to Tomita and Tajima
 \cite{ToTa}. They were the first to workout the second-order perturbation
analysis of the Schwarschild solution with the aim of probing the non-linear
stability of the horizon.

Since the earliest studies, perturbation techniques have been useful
in probing issues of astrophysical significance in situations without
symmetries, that would otherwise be prohibitively complicated to
analyze. Examples of these were the analyses of the motion of
particles in black hole backgrounds.  The formalism was laid out by
Zerilli \cite{zerilli2} and the first studies done by Davis et al.
\cite{dripp}. These have subsequently been extended to rotating holes
and spinning particles (see \cite{Poisson} and references within).

The perturbations of the ``exterior'' Schwarzschild metric could be
coupled to perturbations of interior solutions, and therefore one
could analyze perturbations of stellar objects. This is a subject that
lies outside the scope of this review. For further references see the
work of Cunningham et al.  \cite{cpm} and for more recent references
see  \cite{ToSaMa}. Also outside the scope of this review will be the
treatment of perturbations of rotating black holes, first treated by
Teukolsky \cite{Teuk}, which very recently has been extended to 
second order perturbations \cite{CaLo}.

The increased activity in numerical relativity applied to black hole
collisions  \cite{grandch}, and the possibility of black hole collisions
as a source of detectable gravitational waves  \cite{kipreview}, has
renewed interest in calculations based on perturbation theory. Of
particular interest have been the ``close limit''
calculations  \cite{PrPu}-- \cite{gnppspin}. In this technique, the
colliding holes start out well inside a single horizon. The initial
data, and the spacetime that evolves, outside the horizon, are
considered to be perturbations of a final stationary hole formed in
the collision.  Much of this recent work has been useful in providing
checks of numerical relativity results, providing insight into those
results, and in cautiously extending some results.  The work, so far,
has mostly focused on perturbations of Schwarzschild spacetimes,
although work has very recently started on perturbations of Kerr
holes  \cite{kerrwork}.

An inherent feature of linearized perturbation theory, for this or any
application, is that there is no ``built in'' indication of how good
the perturbation approximation is. For close limit work, linearized
perturbation theory applies, in principle, only in the limit of zero
separation. But in applying this technique, one is not interested so
much in points of principle as in practical results. One wants to know
that a perturbation calculation should agree with ``perfect''
numerical relativity results to, say, within 5\%. There are rough
indicators of when close limit perturbation theory should be
applicable.  For example, if an apparent horizon surrounds two
colliding holes one guesses that close limit evolution of the initial
data will give tolerable accuracy. In other contexts, there are
often similar indicators of when perturbation theory applies. But
these are rough indicators only, and are typically not quantitative.
Short of numerical relativity, the only systematic approach to
quantifying the errors in linearized theory is the study of higher
order perturbations. If one computes a physical quantity correctly to
second order in a perturbation parameter, then the difference between
that result, and the result of first order theory alone, is a
systematic, quantitative, indication of the error in the perturbation
theory calculation.

The need for a good error indicator is paramount in situations where
one is ``pushing the envelope'' of perturbation theory, i.e. venturing
into large values of the expansion parameter. The case of the ``close
limit'' of black hole collisions is one of the first such
situations. One is not really interested in the ``close limit'' of the
collision, one is forced into it since it is tractable, one is really
interested in going as far away as possible from that limit.  For that
reason a significant effort has recently gone into the development of
second order perturbations of the Schwarzschild spacetime as a
practical tool for calculations. We have reported results of second
order calculations for head-on collisions of equal mass
holes  \cite{gnppprl,gnppcqg}, for initially boosted
holes  \cite{gnppboost}, and for a single slowly spinning hole as it
relaxes to its Kerr final state  \cite{gnppspin}.  Those papers
emphasized the results and (especially) the comparison with numerical
relativity results. In the present paper we supply a more detailed
description of how second order perturbation theory is carried out. A
presentation like this one is important if other researchers are to
take advantage of the developed tools in other contexts.

As we mentioned before, there has been some work done in the past on
second order perturbations. Tomita and Tajima  \cite{ToTa} studied
second order perturbations in a null formulation, but their interests
were in studying the stability of the horizon, and therefore their
formalism would have to be significantly reworked to address the issue
we are focusing on, outgoing radiation. Cunningham, {\em et
al.} \cite{cpm} studied the second order perturbations due to the
rotation of a star, if one viewed the rotation as a
perturbation. Because of the narrower range of applicability of these
formalisms we will not cover them in this review.

We will, specifically, describe here a scheme for carrying out second
order perturbation calculations of the evolution of an initial value
solution.  We will assume that on a $t=0$ hypersurface we have a
solution of Einstein's initial value equations for a 3-metric
$\gamma_{ij}$ and an extrinsic curvature $K_{ij}$, and that these can
be expanded to second order in a perturbation parameter. We will
assume, then, that we have first and second order perturbative initial
data. How are these data to be evolved forward in time and, in
particular, how is outgoing radiation to be computed?

We will start with a reasonably careful and complete description of
first order perturbation calculations. This may seem unnecessary in
view of the venerable status of first order work. But we will pattern
almost all our second order equations on their first order
equivalents, and it is important that we establish the notation and
``style'' of the first order approach. This will enormously simplify
understanding second order calculations since it turns out that almost
all second order equations are obvious translations of first order
equations except for the very important addition of ``source'' terms
quadratic in first order perturbations.  Patterning the second order
equations so closely on the first will allow us, when we introduce
second order theory in Sec.\,IV, to focus, not on complexity, but on
gauge issues that are somewhat subtle, and that have no first order
equivalent. These gauge issues are potentially confusing but, as will
be made clear in Sec.\,IV, computations of outgoing radiation cannot
be carried out without facing these issues.

The presentation of the first order formalism will be rather
general. Prescriptions are given for computations starting from any
perturbative initial value solution of Einstein's equations, and
ending with the outgoing radiation amplitudes. We do not give
similarly complete and general prescriptions for the second order
calculations.  The reason is the complexity of the source terms that
appear in the second order equations. For a second order perturbation
of a particular multipole index ($\ell,m$), and a particular parity
(even, odd), there will in general be contributions to source terms
from first order perturbations of all multipole indices and from both
parities.  If we were to give general source terms, the meaning of the
calculations would be obscured by the complexity of the expressions,
and the important, but easily overlooked, conceptual issues might be
missed.  There is a further reason to avoid general source term
expressions. Such source terms would consist of infinite series of
products of first order terms along with a set of coefficients in the
series. In principle, a notation system could be introduced to
represent the Clebsch-Gordon-like factors in these coefficients that
arise from the projection of a tensor multipole component from the
product of two tensor multipoles. In practice, such formalism may be
unrelated to the practical way in which source terms are computed. In
examples in which many first order multipoles contribute to each
second order multipole, it seems plausible that direct numerical
methods, rather than formal methods would be used for the computation of
the source terms. To be concrete, for most second-order applications,
the researcher involved will have to construct the appropriate source
terms. Listing pages and pages of expressions in this paper will not
be a viable route to tackle the problem, since it is impractical to
consider all possible situations. The attempt of this paper is to lay
out the general formalism of how to deal with higher order
perturbations and in particular to address the issues of gauge fixing
and of meaningful extraction of gravitational waveforms. The lessons
developed here are general and useful in all cases. The particular
expressions will have to be re-worked.

Clarity, of course, will require examples of explicit expressions. For
that reason we will take the following approach. Our second order
presentation will be restricted to the $\ell=2$, even parity, vacuum
perturbations. Generalization to other second order multipoles is
straightforward, and the equivalent analysis for odd parity is
significantly simpler. Expressions will not be given for general
source terms (source terms for a general system of first order
perturbations) but will be illustrated with examples from close limit
calculations. It turns out that for the initial data sets that have
been used in close limit work, and until recently as a starting point
for most numerical work, the only first order perturbations are the
$\ell=2$ even parity perturbations. The Misner initial value
solution  \cite{misner} is a particularly simple example of this. In
addition to being important, these cases have anomalously simple
source terms.  This special class of examples will be used in Sec.\,IV
to give concrete illustrations of source terms, and more generally of
nonlinear terms.  The reader is cautioned not to mistake these
nonlinear expressions as generally valid, .

The remainder of the paper is organized as follows. Sec.\,IIA starts
with an overview of perturbation theory in general including
questions of reparameterization and gauge transformations at different
orders. The specialization is made in Sec.\,IIB to perturbations of
the Schwarzschild spacetime, and the Regge--Wheeler (RW) notation (not
to be confused with the Regge--Wheeler gauge) is
introduced  \cite{rw}. The RW gauge  \cite{rw} is then introduced, and we
emphasize the important, but seemingly paradoxical, point that
perturbations in the RW gauge can be considered to be gauge invariant
expressions.  First order perturbation theory is given in Sec.\,III,
starting in Sec.\,IIIA, with the presentation of the ``wave
equations,'' that is, the RW equation  \cite{rw} and the Zerilli
equation  \cite{zprl}, that are used to evolve perturbations in
linearized theory.  Sec.\,IIIB discusses the problem of extracting
outgoing wave amplitudes from the solution of the wave equations, and
partially establishes the pattern that will be used for second order
work. The second order wave equation, a ``second order Zerilli
equation,'' is developed in Sec.\,IVA and is shown, as are all the
second order equations, to be similar to first order equations except
for the inclusion of ``source'' terms quadratic in first order
perturbations.  The problem of extracting second order outgoing wave
amplitude, and the gauge issues involved, are taken up in Sec.\,IVB.
Lastly, a summary is given in Sec.\,V, along with a mention of some
related issues.

A word about our notational conventions is in order.  We will use
coordinates $\{t,r,\theta,\phi\}=\{x^0,x^1,x^2,x^3\}$ always to have the
meaning of Schwarzschild coordinates, in the limit of small
perturbations. Greek indices, $\alpha, \beta, \lambda,\tau,
\mu,\nu\ldots$ will refer to spacetime coordinates, and Latin indices
$i,j,k,\ldots$, will be spatial indices on a constant time surface.
Throughout the paper we will use a consistent scheme for describing
properties of our perturbation functions. A tilde ($\tilde{\ }$)
denotes that a perturbation quantity is in a general gauge, while
``RW'' and ``AF,'' as superscripts to the right of a symbol, mean that
the quantity is defined in the Regge--Wheeler  \cite{rw} or
asymptotically flat gauge. Superscripts in parenthesis to the left of
a quantity will indicate whether the quantity is a first or second
order quantity.  Superscripts in parentheses, to the right of a symbol will
(sometimes) be used to denote multipole indices.  A numerical
subscript to the right of a quantity will be used, following the
notation of Regge and Wheeler \cite{rw}, as part of the name of various
perturbation quantities.  To distinguish even and odd parity 
Regge--Wheeler perturbations, we shall always explicitly add a right
superscript ``odd'' to the odd parity perturbations; the omission of
the ``odd'' indicates that the perturbation is even.  Thus, for
example, The quantity $^{(2)}h^{\rm AF\,(3,0)}_1$ indicates the
axisymmetric octupolar ($\ell=3, m=0$) part of the second order
perturbation of the Regge--Wheeler even parity quantity $h_1$, in an
asymptotically flat gauge.  For compatibility with notation in other
papers, and for other purposes, we shall occasionally add further
information in the form of indices and other 
attachments to 
symbols.  This makes for notation that is sometimes very cumbersome
and is often redundant, but we have found this a price well worth
paying. In a mathematical description in which first and second order
quantities, and quantities in different gauges, are used, all with
similar symbols, clarity of meaning of the symbols is crucial. To
reduce the notation  we shall usually omit the
multipole indices.  Of the various indices, these seem to us to be the
ones that are clearest from context.

On general notational choices we shall in follow the conventions of
Misner, Thorne and Wheeler \cite{mtw} and shall use the sign convention
-+++ for the metric, and units in which $c=G=1$. A dot over an index
shall indicate partial differentiation with respect to $t$. A
subscript following a comma will indicate partial differentiation. For
a function of a single argument a prime will indicate differentiation
with respect to that argument.

\section{Perturbation expansions in general}

\subsection*{A. Basic issues}

Our understanding of perturbation theory in general relativity is
based on the idea of $g_{\alpha\beta}(x^\nu;\epsilon)$ a family of
spacetime metrics parameterized by some physical quantity
$\epsilon$. 
For our family of spacetimes $g_{\alpha\beta}(x^\nu;\epsilon)$, we
suppose that, at least in some restricted region of spacetime,
the metric functions can be written as
\begin{equation}\label{taylor} 
g_{\alpha\beta}(x^\nu;\epsilon)={\ }^{(0)}g_{\alpha\beta}(x^\nu)+
\epsilon\, {\ }^{(1)}g_{\alpha\beta}(x^\nu) +
\epsilon^2\, {\ }^{(2)}g_{\alpha\beta}(x^\nu)\cdots .
\end{equation}The
metric ${\ }^{(0)}g_{\alpha\beta}(x^\nu)$ is  the ``background"
metric, some known solution of Einstein's equations. The expanded form
of the metric in (\ref{taylor}) can be substituted in the Einstein
equations, and the resulting set of equations can be expanded to
orders in $\epsilon$. The equations can then in principle be solved
order by order, first for ${\ }^{(1)}g_{\alpha\beta}(x^\nu)$, (first-order
perturbation theory) then for ${\ }^{(2)}g_{\alpha\beta}(x^\nu)$, and so
forth.

There is an alternative way of viewing first order  perturbation
theory.  One
supposes that there is a {\em particular} spacetime 
$g^{\rm part}_{\alpha\beta}(x^\nu)$ that is in some sense close to a known
``background'' solution ${\ }^{(0)}g_{\alpha\beta}$ of Einstein's
equations. One writes
\begin{equation}\label{LLway}
g_{\alpha\beta}(x^\nu;\epsilon)\equiv
{\ }^{(0)}g_{\alpha\beta}(x^\nu)+\epsilon\left[ g^{\rm
part}_{\alpha\beta}(x^\nu)-{\ }^{(0)}g_{\alpha\beta}(x^\nu)\right]\ .
\end{equation}
and substitutes the right hand side of (\ref{LLway}) into Einstein's
equations, and treats $\epsilon$ as a formal parameter for keeping
track of orders of perturbations. Finally, one sets $\epsilon$ to
unity. This method turns out to give the correct equations for 
treating
$g^{\rm part}_{\alpha\beta}(x^\nu)$
as a first order perturbation of 
${\ }^{(0)}g_{\alpha\beta}$, 
but it is not based on  a systematic 
approach to perturbation theory. For higher order computations, such
a systematic approach is important.

A very relevant example of a family of spacetimes is the
Misner spacetimes, representing two equal mass, initially stationary
throats. The 3-geometry for this initial value solution is given by
\begin{equation}\label{mismet}
ds^2=a^2\widehat{\Phi}^4 \left[d\mu^2+d\eta^2+\sin^2{\eta}\,d\varphi^2
\right]
\end{equation}
where 
\begin{equation}
\widehat{\Phi}\equiv\sum_{n=-\infty}^{+\infty}\frac{1}{
\sqrt{\cosh(\mu+2n\mu_0)-\cos(\eta)}
}\ .
\end{equation}
We now introduce new coordinates $R,\theta,\phi$, related to 
$\mu,\eta,\phi$ in the same way that spherical polar coordinates
are related to bispherical coordinates in flat space:
\begin{equation}                                                
\cosh^2{\mu}=\frac{\left(R^2+a^2\right)^2
}{\left(R^2+a^2\right)^2-\left(2aR\cos{\theta}\right)^2
}\ \ \ \ 
\cos^2{\eta}=\frac{\left(R^2-a^2\right)^2
}{\left(R^2-a^2\right)^2+\left(2aR\sin{\theta}\right)^2
}\ .
\end{equation}
The metric as presented appears in ``isotropic" (conformally flat)
form. In order to recover in the close limit the usual form of the 
Schwarzschild solution we 
transform to a new radial coordinate $r$ through the usual transformation,
\begin{equation}\label{rtrans}
R=\frac{1}{4}\left(\sqrt{r}+\sqrt{r-2M}\right)^2\ ,
\end{equation}
where following Misner we have defined  $M\equiv 4a\Sigma_1$ where
\begin{equation}                                   
\Sigma_1\equiv\sum_{n=1}^\infty\frac{1}{\sinh{n\mu_0}}\ .
\end{equation}

With this new coordinate the 3-geometry takes the form
\begin{equation}\label{M3}                                   
ds^2_{\rm Misner}={\cal F}(r,\theta)^4\left(\frac{dr^2}{1-2M/r}
+r^2d\Omega^2\right)\ ,
\end{equation}                                   
with
\begin{equation}\label{Fdef}                                  
{\cal F}\equiv1+2\,(1+\frac{M}{2R})^{-1}\sum_{\ell=2,4...}
\kappa_\ell(\mu_0)\left(M/R\right)^{\ell+1}P_\ell(\cos\theta)\ ,
\end{equation}                                   
and with the $\kappa_\ell$ coefficients  given by
\begin{equation} \label{kappaeq}                                  
\kappa_\ell(\mu_0)\equiv
\frac{1}{\left(4\Sigma_1\right)^{\ell+1}}\sum_{n=1}^\infty
\frac{(\coth{n\mu_0})^\ell}{\sinh{n\mu_0}}\ .
\end{equation}

In the limit $\mu_0\rightarrow0$, all $\kappa_\ell$ coefficients
vanish, so the Misner initial geometry (more properly, that part of it
covered by $r>2M$) approaches the 3-geometry of a constant $t$ slice
of the Schwarzschild spacetime.  But parameterization for perturbation
theory of the Misner geometry is nontrivial. The ``obvious'' choice of
parameter $\mu_0$ cannot be used, because the deviations from the
Schwarzschild geometry are not linear in $\mu_0$ at small $\mu_0$. The
spacetime {\em can} be expanded in the parameter
$\delta\equiv1/\ln(\mu_0)$. The spacetime will then have nonzero
perturbations at order $\delta^3, \delta^5, \delta^6, \delta^7,
\delta^8,\cdots$. If, however, we project out only the quadrupole
parts of the deviations from the Schwarzschild geometry, the leading
terms in the expansion are of order $\delta^3, \delta^6, \delta^8,
\delta^9,\cdots$.  For consideration of the two lowest orders in this
expansion it is useful to define $\epsilon\equiv\kappa_2(\mu_0)={\cal
  O}(\delta^3)$, and to treat the problem as if we were doing
perturbation to first and second order in $\epsilon$.

\subsubsection*{1. Reparameterization}

The expansion in (\ref{taylor}) is complicated by several issues of
choice.  One of these is simply the question of the choice of
parameter \cite{AbPr}. One could choose a new parameter by $\epsilon= {\cal
E}[\epsilon']$, and treat $\epsilon'$ as the parameter, so that
\begin{equation}\label{taylorprm}
g_{\alpha\beta}(x^\nu;{\cal
E}[\epsilon'])={\ }^{(0)}g_{\alpha\beta}(x^\nu)+
\epsilon'{\ }^{(1)}g'_{\alpha\beta}(x^\nu) + \epsilon'^2\,
{\ }^{(2)}g'_{\alpha\beta}(x^\nu)\cdots\ .
\end{equation}
The condition for the spacetime to have a Taylor expansion in both
$\epsilon$ and $\epsilon'$, is that $\epsilon$ can be expanded as

\begin{equation}\label{reparam} \epsilon=A\epsilon'+B\epsilon'^2+\cdots.  
\end{equation} 
For clarity of explanation, here we will restrict ourselves to
transformations with $A=1$. (This involves no real loss of generality,
since it can always be accomplished with a trivial multiplicative
rescaling.)  With this restriction we have
${\ }^{(1)}g'_{\alpha\beta}(x^\nu)={\ }^{(1)}g_{\alpha\beta}(x^\nu)$. The
implication is that the first-order prediction for a given numerical
value of $\epsilon$ (say $\epsilon=0.1$) is identical to the first
order prediction for the same value of $\epsilon'$ (say
$\epsilon'=0.1$).  Since we are free to choose the function ${\cal
E}$, let us suppose that we choose it such that ${\cal E}[0.1]=10$, say
from the reparameterization $\epsilon=\epsilon'/(1+9.9\epsilon')$.
This means that $\epsilon=0.1$ and $\epsilon'=10$ correspond to the same
physical situation, say the same initial separation between two
coalescing holes.  In the limit as $\epsilon$ gets very small, of course,
the nature of the transformation in (\ref{reparam}) guarantees that
the first-order perturbation method answers will become insensitive to
the choice of parameterization. But that fact may be misleading, since
in practice one does perturbation calculations for specific physical
situations, not as a limit. Let us suppose that $\epsilon=0.1$
represents a physical situation and a parameterization for which
first order theory gives good results. Then our example of a
transformation to $\epsilon'=10$, leads to inaccurate results despite
the fact that the underlying physical problem was amenable to a
perturbative solution. This illustrates the important point that the full
nonlinear nature of a choice, like that of parameterization, can 
have an important effect on the results of first order calculations.

\subsubsection*{2. Gauge transformations}

Special attention is paid to perturbative coordinate transformations,
that is, to transformations that can be written as 
\begin{equation}\label{perttrans}
x'^{\mu}=x^{\mu}+\epsilon{\ }^{(1)}\xi^{\mu}+\epsilon^2{\ }^{(2)}\xi^{\mu}
+\cdots.
\end{equation}
where ${\ }^{(1)}\xi^{\mu}, {\ }^{(2)}\xi^{\mu} ,\cdots$ are functions
of $x^\mu$.  The
transformation in (\ref{perttrans}) induces a transformation of
tensor fields. If (suppressing all indices) we let $T$
represent any tensor then the first order transformation is given by
\begin{equation}\label{lie1}
T'=T-{\cal L}_{\vec{{\ }^{(1)}\xi}}T\ .
\end{equation}  
where $ {\cal L}_{\vec{{\ }^{(1)}\xi}}$ indicates the Lie
derivative \cite{lie} taken with respect to the vector $
\vec{{\ }^{(1)}\xi}$. 
It is well known that such ``gauge'' transformations play an important
role in first order perturbation calculations. Physical answers cannot
depend on coordinate choices, so the coordinate freedom (gauge
freedom) inherent in the gauge transformations must not affect
physical results.  In practice, this can be dealt with by constructing
quantities which are invariant under gauge transformations, and then
using Einstein's equations to compute these quantities. More
typically, the problem is handled by ``gauge fixing,'' i.e., by
imposing specific restrictions on ${\ }^{(1)}g_{\alpha\beta}$ thereby
fixing the coordinate system to first order.

For higher order perturbation computations the basic ideas are the
same but there are crucial differences in detail. First, we remark
that the functions ${\ }^{(2)}\xi^\mu$ in (\ref{perttrans}) are not
vector fields, so the transformation induced on tensor fields cannot
directly be represented by a geometric operation like that in
(\ref{lie1}). In our second order perturbation computations we deal
with this by using a two-step process for gauge transformations.

In the first step the transformation $
x^{\mu'}=x^\mu+\epsilon{\ }^{(1)}\xi^\mu$ is
made, and is carried out to (at least) second order. For a vector
field $V^\mu$, for example, the first step is the evaluation of the
transformed components by
\begin{equation}
V^{\mu'}(x^\alpha)=
\left(\delta^\mu_\alpha+\epsilon{\ }^{(1)}\xi^\mu_{,\alpha}
\right)
V^\alpha(x^{\nu'}-\epsilon {\ }^{(1)}\xi^\nu)\ .
\end{equation}
The terms on the right can be evaluated exactly, but since we shall
end by throwing away everything of order $\epsilon^3$ or higher, we need 
only keep terms of order $\epsilon^2$. If only these terms are kept the
result is
\begin{equation}
\delta V^\alpha\equiv V^{\alpha'}-V^{\alpha}=-\epsilon \left({\cal
L}_{\vec{{\ }^{(1)}\xi}}V\right)^\alpha
+\textstyle{\frac{1}{2}}\epsilon^2 {\ }^{(1)}\xi^\mu{\ }^{(1)}\xi^\nu
V^\alpha_{,\mu\nu} -\epsilon^2{\ }^{(1)}\xi^\nu{\ }^{(1)}\xi^\mu_\alpha
V^\alpha_{,\nu}\ .
\end{equation}
It is of no practical consequence that the $\epsilon^2$
terms do not have a compact geometrical form. It is possible, however, to 
achieve a geometrical form by replacing our simple coordinate transformation
$ x^{\mu'}=x^\mu+\epsilon{\ }^{(1)}\xi^\mu$ by
\begin{equation}\label{brunixtrans}
 x^{\mu'}=x^\mu+\epsilon{\ }^{(1)}\xi^\mu-\textstyle{\frac{1}{2}}
\epsilon^2{\ }^{(1)}\xi^\sigma{\ }^{(1)}\xi^\mu_{,\sigma}\ .
\end{equation}
We find that, to second order in $\epsilon$, the induced transformation
for any tensor field $V$ is
\begin{equation}
\delta V=-\epsilon\left({\cal L}_{\vec{{\ }^{(1)}\xi}}V\right)
+\textstyle{\frac{1}{2}}\epsilon^2
{\cal L}_{\vec{{\ }^{(1)}\xi}}
\left({\cal L}_{\vec{{\ }^{(1)}\xi}}V\right)\ .
\end{equation}
We shall not follow this path here; because of our two step procedure
it really makes little difference. What is important in the first step
is only that all tensor quantities are transformed correctly to
second order in $\epsilon$. 
If the first order transformations were not carried out correctly to 
order $\epsilon^2$, then our tensor fields would change to that order.
The spacetime geometry, in particular, would not be isometric to the
original spacetime geometry. Though it is important that our first
order transformation be correct to second order, the nature of the
second order coordinate change, whether it is that of
(\ref{brunixtrans}), or the first two terms on the right in
(\ref{perttrans}), is irrelevant. The second step will obviate any
specific choice.

We call step 1 the ``first order gauge transformation'' (though it
must be carried out correctly to at least second order). The purpose
of this step will typically be to impose some ``first order gauge
condition,'' that is, some conditions on the first order metric
perturbations ${\ }^{(1)}g_{\alpha\beta}$ that can be brought about by the
proper choice of ${\ }^{(1)}\xi^{\mu}$.

The second step in our procedure is to use  a coordinate 
transformation of the form
\begin{equation}
 x^{\mu'}=x^\mu+\epsilon^2{\ }^{(2)}\xi^\mu\ ,
\end{equation}
and to transform only to lowest order, i.e., to transform tensors 
by
\begin{equation}
\delta V=-\epsilon^2\left({\cal L}_{\vec{{\ }^{(2)}\xi}}V\right)\ .
\end{equation}
This transformation does not change the first order parts of tensor
fields, and hence leaves intact the first order gauge conditions
imposed by the transformation in our first step. With the second step
we can use the choice of the fields ${\ }^{(2)}\xi$ to impose
conditions on ${\ }^{(2)}g_{\alpha\beta}$.  Note that the details of
the functions ${\ }^{(2)}\xi^\mu$ needed to impose these second order
gauge restrictions will depend on the second order part of the metric
{\em after} the first step is performed. The metric, to second order is
affected by the first step. This means that the details of the second
step depend on the details of the first step, and this is why we do not
need to be specific about the second order part of the coordinate
transformation used in the first step.

Our two-step procedure in which we first make the transformation $
x^{\mu'}=x^\mu+\epsilon{\ }^{(1)}\xi^\mu$ and then $
x^{\mu'}=x^\mu+\epsilon^2{\ }^{(2)}\xi^\mu$ leads to the following
explicit transformations of the metric perturbations.

\begin{equation}\label{gauge1}
{\ }^{(1)}g'_{\mu\nu}  = {\ }^{(1)}g_{\mu\nu} -{\ }^{(0)} g_{\mu
\nu},_{\rho}{\ }^{(1)} \xi^{\rho} -
 {\ }^{(0)}g_{\mu \rho} {\ }^{(1)}\xi^{\rho},_{\nu} -
 {\ }^{(0)}g_{\rho \nu} {\ }^{(1)}\xi^{\rho},_{\mu}
\end{equation}
\noindent and,

\begin{eqnarray}\label{gauge2}
{\ }^{(2)}g'_{\mu\nu} & = &
{\ }^{(2)}g_{\mu\nu} \nonumber \\
& &  
 -{\ }^{(1)}g'_{\mu
\nu},_{\rho}{\ }^{(1)} \xi^{\rho} - {\ }^{(1)}g'_{\mu \rho}
{\ }^{(1)}\xi^{\rho},_{\nu} - {\ }^{(1)}g'_{\rho \nu}
{\ }^{(1)}\xi^{\rho},_{\mu} \nonumber \\ & & -{\ }^{(0)} g_{\mu
\nu},_{\rho} {\ }^{(2)}\xi^{\rho} -  {\ }^{(0)}g_{\mu \rho}
{\ }^{(2)}\xi^{\rho},_{\nu} -  {\ }^{(0)}g_{\rho \nu}
{\ }^{(2)}\xi^{\rho},_{\mu} \\
& & - 
{1\over2} {\ }^{(0)}g_{\mu \nu},_{\sigma},_{\lambda}
{\ }^{(1)}\xi^{\sigma}
{\ }^{(1)}\xi^{\lambda}
-  {\ }^{(0)}g_{\mu \lambda},_{\sigma} {\ }^{(1)}\xi^{\sigma}
{\ }^{(1)}\xi^{\lambda},_{\nu} 
\nonumber\\
&&-  {\ }^{(0)}g_{\lambda \nu},_{\sigma}
{\ }^{(1)}\xi^{\sigma} {\ }^{(1)}\xi^{\lambda},_{\mu}
 -  {\ }^{(0)}g_{\sigma \lambda}
{\ }^{(1)}\xi^{\sigma},_{\mu} {\ }^{(1)}\xi^{\lambda},_{\nu}\ .
\nonumber
\end{eqnarray}

Throughout this paper we shall use the two step procedure for second
order gauge transformations. An alternative treatment has recently
been presented by Bruni {\em et al.} \cite{bruni}. That treatment gives a
much more geometrical view of second order gauge transformations.

\subsubsection*{4. Standard hierarchy}

For definiteness we shall now consider specifically the vacuum
Einstein equations, which is in fact the case of interest for the
close-limit method. Einstein's vacuum equations can be written as
${\cal G}_{\lambda\tau}
\left(g_{\alpha\beta}(x^\nu)\right)=0$, where
 ${\cal G}_{\lambda\tau}$ represents the actions of combining zeroth,
first and second derivatives of $g_{\alpha\beta}$ to form the
component $G_{\lambda\tau}$ of the Einstein tensor. If the expansion
in (\ref{taylor}) is used, the Einstein equations for the family of
solutions
\begin{equation}\label{Gexpan}
{\cal G}_{\lambda\tau}\left(g_{\alpha\beta}(x^\nu;\epsilon)
\right)
=
{\cal G}_{\lambda\tau}\left(
{\ }^{(0)}g_{\alpha\beta}(x^\nu)+
\epsilon {\ }^{(1)}g_{\alpha\beta}(x^\nu) +
\epsilon^2 {\ }^{(2)}g_{\alpha\beta}(x^\nu)\cdots 
\right)=0\ ,
\end{equation}
can be expanded in powers of $\epsilon$. The terms to order 0 in
$\epsilon$ are just those of ${\cal
G}\left({\ }^{(0)}g_{\alpha\beta}(x^\nu)\right)=0 $, and are satisfied
immediately since, by assumption, ${\ }^{(0)}g_{\alpha\beta}$ is a
solution to Einstein's equations.

The terms in 
(\ref{Gexpan})
of first order in $\epsilon$
can be written in the form 
\begin{equation}\label{first}
\epsilon L_{\lambda\tau}({\ }^{(1)}g_{\alpha\beta})=0\ ,
\end{equation}
where $L_{\lambda\tau}$, formally defined by
\begin{equation}
L_{\lambda\tau}\equiv\left.\frac{\partial}{\partial\epsilon}{\cal G}_{\lambda\tau}\left(g_{\alpha\beta}(x^\nu;\epsilon)
\right)
\right|_{\epsilon=0}
\end{equation}
is a linear operator on $ {\ }^{(1)}g_{\alpha\beta} $,
consisting of combinations of differentiation and multiplications by
specific coordinate functions. The details of $L$ depend on the
background solution ${\ }^{(0)}g_{\alpha\beta}$.  The equations for
${\ }^{(1)}g_{\alpha\beta}$ contained in (\ref{first}), constitute first
order perturbation theory. 

The part of (\ref{Gexpan}) that is
proportional to $\epsilon^2$ has terms of two different types. There will
be terms linear in ${\ }^{(2)}g_{\alpha\beta}$ and terms quadratic in
${\ }^{(1)}g_{\alpha\beta}$. The terms of the first type occur in
precisely the same way as do the ${\ }^{(1)}g_{\alpha\beta}$ terms in the
first order expression. We can therefore write the second order part of
(\ref{Gexpan}) as
\begin{equation}\label{second}
\epsilon^2 L_{\lambda\tau}({\ }^{(2)}g_{\alpha\beta})=
\epsilon^2{\ }^{(2)}{\cal T}_{\lambda\tau}\left(
{\ }^{(1)}g_{\alpha\beta}
\right)\ ,
\end{equation}
where ${\ }^{(2)}{\cal T}_{\lambda\tau}$
is quadratic in ${\ }^{(1)}g_{\alpha\beta}$.  One views
(\ref{second}) as a set of linear equations for
${\ }^{(2)}g_{\alpha\beta}$, with the right hand side a ``source''
which
is known from the solution of the first order problem.
In a similar manner one goes on to find that terms in 
(\ref{Gexpan}) higher order in $\epsilon$ have the form
\begin{equation}\label{gen}
\epsilon^n 
L_{\lambda\tau}^{(n)}(g_{\alpha\beta})=\epsilon^n
{\ }^{(n)}{\cal T}_{\lambda\tau}
\left(
{\ }^{(n-1)}g_{\alpha\beta},{\ }^{(n-2)}g_{\alpha\beta},{\
}^{(n-3)}g_{\alpha\beta},\cdots
\right)\ .
\end{equation}
In the $n^{\rm th}$ order source term ${\ }^{(n)}{\cal
T}_{\lambda\tau}$ the combinations of the lower order metric
perturbations must occur according to obvious rules. For example, the
source ${\ }^{(5)}{\cal T}_{\lambda\tau}$, for the fifth order
perturbations, will  in general have contributions including terms
of fifth power in ${\ }^{(1)}g_{\alpha\beta}$, terms of second power
in ${\ }^{(1)}g_{\alpha\beta}$ multiplied by terms linear in ${\
}^{(3)}g_{\alpha\beta}$, and so forth. (Here ``terms in ${\
}^{(k)}g_{\alpha\beta}$'' includes derivatives of these terms.)  In
principle, one can solve order by order since the source terms for
each order are given by the solution known from the lower
orders. Furthermore, at each order the linear operator being solved is
precisely the same $L_{\lambda\tau}$; all that is changing is the
source terms and the initial conditions.

There are alternatives to this ``standard hierarchy'' of equations for
perturbations of increasing order.  One could, for example, adopt the
following alternative iterative scheme.  Start with the equation
\begin{equation}\label{alternate}
{\cal G}\left(
{\ }^{(0)}g_{\alpha\beta}(x^\nu)+
\epsilon {\ }^{(1)}g_{\alpha\beta}(x^\nu) +
\epsilon^2 {\ }^{(2)}g_{\alpha\beta}(x^\nu)\cdots 
\right)\ =0.
\end{equation}
As a first step, keep only the terms that are first order in
 $\epsilon$. This will give a set of linear equations for
 ${\ }^{(1)}g_{\alpha\beta}$ identical to (\ref{first}).  
Next, in (\ref{alternate}), substitute  the known solution for
${\ }^{(1)}g_{\alpha\beta}$, omit ${\ }^{(k)}g_{\alpha\beta}$ terms for
$k>2$ and solve the resulting equations to lowest order in
${\ }^{(2)}g_{\alpha\beta}$.  The resulting linear equations for
${\ }^{(2)}g_{\alpha\beta}$ will {\em not} be the same as those in
(\ref{second}). In the present method we are in effect feeding back
{\em all } the information about the metric resulting from the first
order solution, and we are therefore keeping different terms that are
higher oder in $\epsilon$.  In this alternative method, the
differential equation for ${\ }^{(2)}g_{\alpha\beta}$ represents
propagation of the second order perturbations on a background spacetime
correct to first order. In the standard hierarchy the second order
perturbations, and all perturbations, propagate on the zero order
background.

This alternative method would seem intuitively to offer advantages,
and indeed is being investigated, but it entails a very serious
practical difficulty. In the standard hierarchy, the differential
operator $L$ embodies the simplicity and symmetries of the background
solution.  In practice this means, for example, that multipole
decomposition for angular variables can be used.  In the alternative
method the differential operator for higher order perturbations no
longer has those simplifications. The higher order equations, as in
the standard hierarchy, are linear, but they would need to be solved
numerically as 2+1 or 3+1 linear hyperbolic systems.

\subsection*{B. Schwarzschild perturbations} 
We specialize now to a spherically symmetric background and introduce
coordinates $\{x^0 , x^1 , x^2 ,x^3\} = \{t, r, \theta, \phi\}$ chosen
such that in the $\epsilon\rightarrow0$ limit $\theta, \phi$ become the
usual spherical coordinates.  The ``background operator'' $L$ acting on
the unknown perturbations embodies the spherical symmetry of the
background, and so allows us to eliminate angular variables $\theta,
\phi$.  To take advantage of the symmetry it is necessary to expand the
metric
perturbations in tensor spherical harmonics. This does not require
that we make further assumptions about the background geometry, but
below we shall specialize to the case of a Schwarzschild geometry. To
avoid the repetition of very similar lengthy expressions, we give here
the description of the multipole decomposition specific to the
Schwarzschild background, where we follow the notations and
conventions \cite{rw} of Regge and Wheeler (RW).

Our background
metric ${\ }^{(0)}g_{\mu \nu}$ is the standard exterior Schwarzschild
solution in Schwarzschild's coordinates
%{}^{(0)}g_{\mu \nu} = 
\begin{equation}
a=
\begin{array}{cccc}
-(1-2M/r)& 0 & 0  & 0\cr
0 & (1-2M/r)^{-1} & 0 &  0  \\ 
0  &  0  &     r^2 & 0 \\ 
0 & 0 &0 & r^2 \sin^2{\theta} 
\end{array}
\ .
\end{equation}
The 10 metric perturbations can be divided into two sets, called
perturbations of odd and even parity, which are not mixed by
tensorial operators which respect spherical symmetry. The multipole
decomposition, is given below for even parity perturbations
%Y_\ell^{m}
\begin{eqnarray}
{\ }^{(n)}g_{0 0} & = & \left(1 - {2M \over r} \right)\sum_{\ell,m}{\
}^{(n)}\widetilde{H_0}^{(\ell,m)} Y_\ell^{m}\nonumber \\ {\
}^{(n)}g_{0 1} & = & \sum_{\ell,m}{\ }^{(n)}\widetilde{H_1}^{(\ell,m)} 
Y_\ell^{m}
\nonumber \\ {\ }^{(n)}g_{0 2} & = & \sum_{\ell,m}{\
}^{(n)}\widetilde{h_0}^{(\ell,m)} {\partial Y_\ell^{m}\over \partial
\theta}\nonumber \\ {\ }^{(n)}g_{0 3} & = & \sum_{\ell,m} {\
}^{(n)}\widetilde{h_0}^{(\ell,m)} {\partial Y_\ell^{m} \over \partial
\phi}\nonumber \\ {\ }^{(n)}g_{1 1} & = & \sum_{\ell,m}\left(1 - {2M
\over r} \right)^{-1} {\ }^{(n)}\widetilde{H_2}^{(\ell,m)} Y_\ell^{m}
\nonumber \\ {\ }^{(n)}g_{1 2} & = & \sum_{\ell,m}
{\ }^{(n)}\widetilde{h_1}^{(\ell,m)} {\partial Y_\ell^{m} \over
\partial \theta}\label{startpertdef} \\ {\ }^{(n)}g_{1 3} & = &
\sum_{\ell,m} {\ }^{(n)}\widetilde{h_1}^{(\ell,m)} {\partial Y_\ell^{m}
\over \partial \phi}\nonumber \\ {\ }^{(n)}g_{2 2} & = &
r^2\sum_{\ell,m} \left( {\ }^{(n)}\widetilde{K}^{(\ell,m)} +
{\ }^{(n)}G^{(\ell,m)} {\partial^ 2
\over
\partial \theta^2} \right) Y_\ell^{m} \nonumber \\ {\ }^{(n)}g_{2 3} &
= & r^2\sum_{\ell,m} {\ }^{(n)}\widetilde{G}^{(\ell,m)} \left(
{\partial^2
\over \partial \theta \partial \phi} - 
{\cos \theta \over \sin \theta} {\partial \over 
\partial \phi} \right) Y_\ell^{m} \nonumber \\ {\ }^{(n)}g_{3 3} & = &
r^2 \sum_{\ell,m}\left[ {\ }^{(n)}\widetilde{K}^{(\ell,m)}  \sin^2
(\theta) + {\ }^{(n)}\widetilde{G}^{(\ell,m)}   \left({\partial^2
\over \partial \phi^2} + \sin \theta \cos \theta  {\partial \over
\partial \theta} \right) \right] Y_\ell^{m} \nonumber\ ,
\end{eqnarray}
and odd parity perturbations
\begin{eqnarray}
{\ }^{(n)}g_{0 2\rm odd} & = &-\sum_{\ell,m}\frac{{\
}^{(n)}\widetilde{h_0}^{(\ell,m)\rm odd}}{\sin\theta}\frac{\partial
Y_\ell^{m}}{\partial\phi}\nonumber\\
{\ }^{(n)}g_{03\rm odd} & = &\sum_{\ell,m}{\
}^{(n)}\widetilde{h_0}^{(\ell,m)\rm odd}\sin\theta\frac{\partial
Y_\ell^{m}}{\partial\theta}\nonumber\\
{\ }^{(n)}g_{1 2\rm odd} & = &-\sum_{\ell,m}\frac{{\
}^{(n)}\widetilde{h_1}^{(\ell,m)\rm odd}}{\sin\theta}\frac{\partial
Y_\ell^{m}}{\partial\phi}\nonumber\\ 
{\ }^{(n)}g_{1 3\rm odd} & = &\sum_{\ell,m}{\
}^{(n)}\widetilde{h_1}^{(\ell,m)\rm odd}\sin\theta\frac{\partial
Y_\ell^{m}}{\partial\theta}
\label{endpertdef}\\
{\ }^{(n)}g_{2 2\rm odd} & = &\sum_{\ell,m}{\
}^{(n)}\widetilde{h_2}^{(\ell,m)\rm odd}
\left(\frac{1}{\sin\theta}\frac{\partial^2}{\partial\theta\partial\phi}
-\frac{\cos\theta}{\sin^2\theta}\frac{\partial}{\partial\phi}
\right)
Y_\ell^{m}\nonumber
\\
{\ }^{(n)}g_{3 3\rm odd} & = &-\sum_{\ell,m}{\
}^{(n)}\widetilde{h_2}^{(\ell,m)\rm odd}
\left(\sin\theta\frac{\partial^2}{\partial\theta\partial\phi}
-\cos\theta\frac{\partial}{\partial\phi}
\right)
Y_\ell^{m}
\nonumber
\\
{\ }^{(n)}g_{2 3\rm odd} & =
&\frac{1}{2}\sum_{\ell,m}{\ }^{(n)}\widetilde{h_2}^{(\ell,m)\rm odd}
\left(
\frac{1}{\sin\theta}\frac{\partial^2}{\partial\phi^2}
+\cos\theta\frac{\partial}{\partial\theta}-\sin\theta
\frac{\partial^2}{\partial\theta^2}
\right)Y_\ell^{m}\nonumber\ ,
\end{eqnarray}
where $\widetilde{H_0}^{\ell,m)}$, $\widetilde{H_1}^{(\ell,m)}$,
$\widetilde{H_2}^{(\ell,m)}$, $\widetilde{h_0}^{(\ell,m)}$,
$\widetilde{h_1}^{(\ell,m)}$, $\widetilde{K}^{(\ell,m)}$ and
$\widetilde{G}^{(\ell,m)}$, are functions of $r$ and $t$ for the even
parity parts of the perturbations, and ${\
}^{(n)}\widetilde{h}^{(\ell,m,\rm odd)}_{0}$, ${\
}^{(n)}\widetilde{h}^{(\ell,m,\rm odd)}_{1}$, ${\
}^{(n)}\widetilde{h}^{(\ell,m),\rm odd}_{2}$ are the odd parity
functions. The complete metric perturbations are the sum of the even
and the odd parity parts. Here we introduce the notation that a tilde
$\widetilde{\ }$ over a perturbation function indicates that it is
in an arbitrary gauge. As an example of metric perturbations we have,
from (\ref{M3}) and 
(\ref{Fdef}), 
the initial perturbations of the Misner spacetime.  For $\ell=2$, these are
\begin{equation}\label{Misexample}
{\ }^{(1)}\widetilde{H}_2={\ }^{(1)}\widetilde{K}=
8\kappa_2\left(
1-\frac{2M}{R}
\right)^{-1} \left(
\frac{2M}{R}
\right)^3\sqrt{\frac{4\pi}{5}}\ ,
\end{equation}
and ${\ }^{(1)}\widetilde{G}={\ }^{(1)}\widetilde{h}_1=0$.
We can choose to set the initial lapse and shift such that 
${\ }^{(1)}\widetilde{H}_0={\ }^{(1)}\widetilde{H}_1
={\ }^{(1)}\widetilde{h}_0=0$.

We now fix coordinates to first order by demanding that the
perturbations satisfy the RW gauge conditions: For even
parity this is the condition that the first order even parity
functions ${\ }^{(1)}h_0^{(\ell,m)}$, $ {\ }^{(1)}h_1^{(\ell,m)}$, $
{\ }^{(1)}G^{(\ell,m)} $ vanish. The odd parity RW
gauge choice is that the function ${\ }^{(1)}h_2^{(\ell,m)\rm odd}$
vanishes. 
This specialization  is accomplished with the first order gauge
transformation (\ref{gauge1}), with the following notation:
\begin{eqnarray}
{\ }^{(1)}\xi^0&=&{\ }^{(1)}A_0 Y_\ell^{m}\label{firstxi}\\
{\ }^{(1)}\xi^1&=&{\ }^{(1)}A_1 Y_\ell^{m}\\
{\ }^{(1)}\xi^2&=&{\ }^{(1)}A_2
Y_\ell^{m},_\theta
+{\ }^{(1)}B^{\rm odd}
Y_\ell^{m},_\phi/\sin{\theta}\\
{\ }^{(1)}\xi^3&=&{\ }^{(1)}A_2
Y_\ell^{m},_\phi
/\sin^2{\theta}-{\ }^{(1)}B^{\rm odd}
Y^m_{\ell},_\theta/\sin{\theta}
\end{eqnarray}
These gauge functions give transformations of the RW metric
perturbation functions as follows

\begin{eqnarray}
{\ }^{(1)}{h}_0^{\rm RW,odd}&=&{\ }^{(1)}\widetilde{h}_0^{\rm odd}+
r^2\frac{\partial{\ }^{(1)}B^{\rm odd}}{\partial t}\nonumber\\
{\ }^{(1)}{h}_1^{\rm RW,odd}&=&{\ }^{(1)}\widetilde{h}_1^{\rm odd}
+r^2\frac{\partial{\ }^{(1)}B^{\rm odd}}{\partial r}\nonumber\\
{\ }^{(1)}{h}_2^{\rm RW,odd}&=&{\ }^{(1)}h_2^{\rm
odd}-2r^2{\ }^{(1)}B^{\rm odd}\nonumber\\ {\ }^{(1)}{H}^{\rm RW}_0 & =
& {\ }^{(1)}\widetilde{H}_0 + {2M \over r(r-2M)}{\ }^{(1)} A_1 +2
{\partial {\ }^{(1)}A_0 \over \partial t}  \nonumber \\
{\ }^{(1)}{H}^{\rm RW}_1 & = & {\ }^{(1)}\widetilde{H}_1 - {r \over
r-2M} {\partial {\ }^{(1)}A_1\over \partial t}  +{r-2M \over r}
{\partial {\ }^{(1)}A_0\over \partial r}  \nonumber \\
{\ }^{(1)}{h}^{\rm RW}_0 & = & {\ }^{(1)}\widetilde{h}_0 - r^2
{\partial {\ }^{(1)}A_2
\over \partial t} +{r-2M \over r} {\ }^{(1)}A_0 \nonumber \\
\label{gauge2p}
{\ }^{(1)}{H}^{\rm RW}_2 & = & {\ }^{(1)}\widetilde{H}_2+ {2M \over r(r-2M)} 
{\
}^{(1)}A_1
-2 {\partial {\ }^{(1)}A_1\over \partial r } \\ {\ }^{(1)}{h}^{\rm
RW}_1 & = & {\ }^{(1)}\widetilde{h}_1 - {r \over r-2M} {\ }^{(1)}A_1 -
r^2 {\partial {\ }^{(1)}A_2\over \partial r} \nonumber \\
{\ }^{(1)}{G}^{\rm RW} & = & {\ }^{(1)}\widetilde{G} - 2 {\ }^{(1)}A_2
\nonumber \\ {\ }^{(1)}{K}^{\rm RW} & = & {\ }^{(1)}\widetilde{K} - {2
\over r}{\
}^{(1)}A_1 \nonumber\ .
\end{eqnarray}
with the  gauge functions
\begin{eqnarray}
{\ }^{(1)}A_0&=&\left(\frac{1}{2}r^2 
\frac{\partial{\ }^{(1)}\widetilde{G}}{\partial
t}-{\ }^{(1)}\widetilde{h_0}\right)\\
{\ }^{(1)}A_1&=&(1-2M/r)\left(-\frac{1}{2}r^2
\frac{\partial{\ }^{(1)}\widetilde{G}}{\partial r}
+{\ }^{(1)}\widetilde{h_1}\right)\\
{\ }^{(1)}A_2&=&\textstyle{\frac{1}{2}}{\ }^{(1)}\widetilde{G}\\
{\ }^{(1)}B^{\rm odd}&=& {1 \over 2 r^2}
{\ }^{(1)}\widetilde{h_2}^{\rm odd}\label{gaugeB}\ .
\end{eqnarray}

With this choice of first order gauge transformation, the result for
the first order perturbations is:
\begin{eqnarray}
{\ }^{(1)}h_0^{\rm RW,odd}&=&\tilde{h}_0^{\rm odd}
+{\textstyle\frac{1}{2}} {\ }^{(1)}\widetilde{h_2}^{\rm
odd},_t\label{oddtrans1}\\
{\ }^{(1)}h_1^{\rm RW,odd}&=&{\ }^{(1)}\tilde{h}_1^{\rm odd}
+{\textstyle\frac{1}{2}}r^2
\left(r^{-2}{\ }^{(1)}\widetilde{h_2}^{\rm odd}
\right),_r\label{oddtrans2}
\end{eqnarray}
and those for the even-parity functions are
\begin{eqnarray}
{\ }^{(1)}K^{RW} & = & {\ }^{(1)}\widetilde{K} +(r-2M) \left(
{\ }^{(1)}\widetilde{G},_r  -{2 \over r^2}{\ }^{(1)}\widetilde{h}_1
\right)\label{eventrans1} \\
{\ }^{(1)}H_2^{RW} & = &
{\ }^{(1)}\widetilde{H}_2 + (2r-3M) \left({\ }^{(1)}\widetilde{G},_r -
{2 \over r^2}{\ }^{(1)}\widetilde{h}_1 \right) + r(r-2M)
\left({\ }^{(1)}\widetilde{G},_r - {2 \over
r^2}{\ }^{(1)}\widetilde{h}_1 \right),_{r}
   \\
{\ }^{(1)}H_1^{RW} & = &
{\ }^{(1)}\widetilde{H}_1+r^2 {\ }^{(1)}\widetilde{G},_{tr} -
{\ }^{(1)}\widetilde{h}_1,_t - {2M \over r(r-2M)}
{\ }^{(1)}\widetilde{h}_0 + {\ }^{(1)}\widetilde{h}_0,_r + {r(r-3M)
\over r-2M} {\ }^{(1)}\widetilde{G},_t
   \label{eventrans3}\\
{\ }^{(1)}H_0^{RW}  & = &  {\ }^{(1)}\widetilde{H}_0 -M
\left({\ }^{(1)}\widetilde{G},_r - {2  \over r^2
}{\ }^{(1)}\widetilde{h}_1 \right) + {2 r \over r-2M
}{\ }^{(1)}\widetilde{h}_0,_t +{r^3 \over
(r-2M)}{\ }^{(1)}\widetilde{G},_{tt}\label{eventrans4}\ .
\end{eqnarray}
Here we have dropped the $\ell,m$ indices in order 
to simplify the intricate notation. 

The above equations show that we can choose to view the left-hand
quantities not as metric perturbations expressed in a particular
coordinate gauge, but rather (due to the right hand side) as
combinations of metric perturbations expressed in an arbitrary
coordinate gauge. In this sense we can, and we will, view the RW-gauge
quantities, such as $H_1^{RW}$ as a compact expression for a
general-gauge expression. It should be emphasized that it is trivial,
combining the above results to construct expressions that are
{\em explicitly gauge invariant} in terms of the metric perturbations in 
an {\em arbitrary} gauge.

It is instructive to point out the somewhat special aspects of the
RW gauge choice and to compare it to other approaches.
Chandrasekhar \cite{chandrabook} chooses to work in a diagonal metric, for 
axisymmetric perturbations,
by making the even parity restrictions ${\ }^{(1)}H_1^{\ell m}={\
}^{(1)}h_0^{\ell m}={\ }^{(1)}h_1^{\ell m}=0$. This choice turns out not
to uniquely fix the gauge \cite{priceipser}. One can, for example, perform
a nontrivial transformation (\ref{gaugeB}) with ${\ }^{(1)}A_0=0$,
with ${\ }^{(1)}A_2$ any function of $r$ alone, and with 
${\ }^{(1)}A_1$ given by
\begin{equation}
{\ }^{(1)}A_1=-r(r-2M)\frac{
\partial{\ }^{(1)}A_2}{\partial r}\ .
\end{equation}
Such a transformation leaves unchanged the values of ${\
}^{(1)}H_1^{\ell m}, {\ }^{(1)}h_0^{\ell m}$ and ${\ }^{(1)}h_1^{\ell
m}$, but changes other perturbations. This means that the metric
perturbations in the Chandrasekhar gauge are not unique, and thus it is
impossible for them to be expressed uniquely in terms of arbitrary
gauge metric perturbations, as the RW perturbations are in
(\ref{eventrans1}) -- (\ref{eventrans4}).

Another approach to perturbation theory is that of
Moncrief \cite{moncrief} who, for analysis of even parity
perturbations, uses gauge invariant combinations only of the
3-geometry quantities ${\ }^{(1)}\widetilde{H}_2^{\ell m}, {\
}^{(1)}\widetilde{K}^{\ell m}, {\ }^{(1)}\widetilde{G}^{\ell m}$ and
${\ }^{(1)\widetilde{h}_1^{\ell m}}$. This approach is quite useful
for connecting the computation of radiation to initial value data,
since the quantities used are, by construction, independent of the
(necessarily arbitrary) lapse and shift. The Moncrief approach will be
discussed again below. Here we confine our attention to the question
of whether one could invoke a ``Moncrief gauge,'' a gauge in which
perturbations of the lapse and shift are set to zero, by choosing ${\
}^{(1)}H_1^{\ell m}={\ }^{(1)}H_0^{\ell m}={\ }^{(1)}h_0^{\ell m}=0$.
In such a gauge computations would only involve perturbations of the 3
geometry. It is easy to show, however, that such a choice suffers from
the same problem as the Chandrasekhar gauge. The gauge conditions do
not completely fix the gauge.

The transformation to the RW gauge (or the derivation of the RW
arbitrary-gauge expressions) has been discussed as a first order
problem.  We now view this as the first step in the two step process we
discussed above.  Our second step is a second order gauge
transformation [see (\ref{perttrans})] of the form
\begin{equation}
x^{\mu'}=x^\mu+\epsilon^2{\ }^{(2)}\xi^\mu .
\end{equation}
We choose ${\ }^{(2)}\xi^\mu$ to impose the RW gauge conditions (e.g.,
${\ }^{(2)}G=0$)
to second order. The form of 
${\ }^{(2)}\xi^\mu$ needed is exactly that of $\xi^\mu$
in (\ref{firstxi}) through (\ref{gaugeB}), with the index ``1''
replaced by ``2''.  Thus for example, second order gauge
transformations can be written as:
\begin{eqnarray}
{\ }^{(2)}\xi^0&=&{\ }^{(2)}A_0Y_\ell^m\label{firstzeta}\\
{\ }^{(2)}\xi^1&=&{\ }^{(2)}A_1Y_\ell^m\\
{\ }^{(2)}\xi^2&=&{\ }^{(2)}A_2Y^m_{\ell},_\theta
+{\ }^{(2)}B^{\rm odd}Y^m_{\ell},_\phi/\sin{\theta}\\
{\ }^{(2)}\xi^3&=&{\ }^{(2)}A_2Y^m_{\ell},_\phi/\sin^2{\theta}
-{\ }^{(2)}B^{\rm odd}Y^m_{\ell},_\theta/\sin{\theta}\ .
\end{eqnarray}
To set ${\ }^{(2)}H_1^{\ell m}={\ }^{(2)}h_0^{\ell m}={\
}^{(2)}h_1^{\ell m}=0$ we use, for example, ${\
}^{(2)}A_2=\textstyle{\frac{1}{2}}{\ }^{(2)}\widetilde{G}$, and find
relations between the second order RW perturbations and the second
order perturbations in a general gauge, for example
\begin{equation}
{\ }^{(2)}K^{RW}  = {\ }^{(2)}\widetilde{K} 
+(r-2M) \left( {\ }^{(2)}\widetilde{G},_r  
-2r^{-2}
{\ }^{(2)}\widetilde{h}_1 \right).
\end{equation}

It should be mentioned that there appear at second order certain
ambiguities in going to the RW gauge if the first order perturbations
have $\ell=0$ components. The gauge fixing procedure leaves a residual
gauge symmetry that has to, and can be, dealt with. This issue has
been discussed in detail in  \cite{gnppboost}, and we will not address
it here.

\section{First order perturbations}

\subsection{Wave  equations}

The first order part of Einstein's equations, the content of 
(\ref{first}),
constitutes linearized perturbation theory. Though this theory is long
established, we review it here since it forms the foundation of our
second order computations. Even and odd parity perturbations completely
decouple in first order theory and we can analyze them separately.  It
is useful to consider first the relatively simple odd parity
problem, first solved  in 1957 by Regge
and Wheeler \cite{rw}.  In the specialized RW gauge 
the nontrivial odd parity equations are 
\begin{eqnarray}
\frac{\partial^2{\ }^{(1)}h_0^{\rm RW}}{\partial r^2}
-\frac{\partial^2{\ }^{(1)}h_1^{\rm RW}}{\partial r\partial t}
-\frac{2}{r}\frac{\partial {\ }^{(1)}h_1^{\rm RW}}{\partial t}
+\left[\frac{4M}{r}-\ell(\ell+1)
\right]\frac{    {\ }^{(1)}h_0^{\rm RW}    }{r(r-2M)}=0\\
\frac{\partial^2{\ }^{(1)}h_1^{\rm RW}}{\partial r^2}
-\frac{\partial^2{\ }^{(1)}h_0^{\rm RW}}{\partial r\partial t}
+(\ell-1)(\ell+2)(r-2M)\frac{    {\ }^{(1)}h_1^{\rm RW}    }{r^3}=0\\
\left(1-\frac{2M}{r}
\right)\frac{\partial{\ }^{(1)}h_1^{\rm RW}}{\partial r}
-\left(1-\frac{2M}{r}
\right)\frac{\partial{\ }^{(1)}h_0^{\rm RW}}{\partial t}
+\frac{2M}{r^2}{\ }^{(1)}h_0^{\rm RW}=0\ .\label{oddEinst1}
\end{eqnarray}
Regge and Wheeler \cite{rw} defined, for each multipole, the wavefunction
\begin{equation}\label{RWQ}
Q^{\rm odd}\equiv r^{-1}(1-2M/r)h_1^{\rm RW, odd}\ .
\end{equation}
The field equations above show that this perturbation quantity
decouples from ${\ }^{(1)}h_0^{\rm RW, odd}$, and satisfies the wave
equation
\begin{equation}\label{rweq}
\frac{\partial^2Q^{\rm odd}}{\partial r^{*2}}
-\frac{\partial^2Q^{\rm odd}}{\partial t^{2}}
+Q^{\rm odd}V_\ell^{\rm RW}(r)=0\ .
\end{equation}
Here $V_\ell^{\rm RW}$ is an $\ell$-dependent function of $r$ that acts
as a potential in the wave equation, and $r^*\equiv r-2M\ln{(r/2M-1)}$
is the ``tortoise'' radial coordinate introduced by Regge and Wheeler.
Once (\ref{rweq}) is solved and $Q^{\rm odd}$, and hence $h_1^{\rm RW,
odd}$ are known, one can find $h_0^{\rm RW, odd}$ by
solving (\ref{oddEinst1}). In solving that equation, the integration
constant is supplied by the specification of $h_0^{\rm RW, odd}$ on
the initial $t=0$ hypersurface (which is related to the initial
extrinsic curvature; see below).  Once $h_1^{\rm RW, odd}$ and
$h_0^{\rm RW, odd}$ are known, all gauge invariant odd-parity
information is known.

The odd-parity wave equation (\ref{rweq}) requires of course the
specification of Cauchy data, and these must be supplied from the
first order perturbations of the 3-metric $\gamma_{ij}$ and the
extrinsic geometry $K_{ij}$ of an initial value solution. Here and
throughout we will take our initial surface to be a surface at $t=0$.
From initial information about the perturbed 3-metric ${\
}^{(1)}\gamma_{ij}$ in some gauge, we can immediately infer the value
of ${\ }^{(1)}\widetilde{h_1}^{\rm odd}$ at $t=0$ and ${\
}^{(1)}\widetilde{h_2}^{\rm odd}$ at $t=0$. With these, and with
(\ref{oddtrans2}), we can compute ${\ }^{(1)}{h_1}^{\rm RW,odd}$ at
$t=0$ and hence can compute $Q^{\rm odd}$.  The general relationship
between the extrinsic curvature and the time derivative of the metric is
\begin{equation}
K_{ij}=\frac{1}{2N}\left[
g_{0i|j}+g_{0j|i}-\frac{\partial}{\partial t}g_{ij}
\right]\ .
\end{equation}
Here $N$ is the lapse function, given by $\sqrt{-1/g^{tt}}$.
The initial value of the odd-parity shift, 
${\ }^{(1)}\widetilde{g}_{t\theta}^{\rm odd}, 
{\ }^{(1)}\widetilde{g}_{t\phi}^{\rm odd}$
or equivalently 
${\ }^{(1)}\tilde{h}_0^{\rm odd}$
can be freely specified, and it is convenient to set 
${\ }^{(1)}\tilde{h}_0^{\rm odd}$
initially to zero. With this choice, we have
\begin{equation}\label{Kgamdot}
K_{ij}=
%=-\sqrt{1-\frac{2M}{r}}\ 
{1 \over 2} 
\sqrt{ r \over r-2M}\
\frac{\partial}{\partial t}g_{ij}.
\end{equation}
The time derivative of ${\ }^{(1)}\tilde{h}_2^{\rm odd}$ follows from
the odd parity part of the $\theta\theta, \phi\phi$, or $\theta\phi$
components of this equation. The time derivative of ${\
}^{(1)}\tilde{h}_1^{\rm odd}$ follows from the odd parity part of the
$r\theta$ or $r\phi$ components. From these time derivatives, and from
(\ref{oddtrans2}), we find the time derivative of 
${\ }^{(1)}{h_1}^{\rm RWodd}$  and hence of $Q^{\rm odd}$.

Moncrief has taken a distinctly different approach. He works in an
arbitrary gauge with only the perturbations of the 3-geometry. (Thus
for odd parity he works with ${\ }^{(1)}\widetilde{h_1}^{\rm odd}$ and
${\
}^{(1)}\widetilde{h_2}^{\rm odd}$, but not ${\
}^{(1)}\widetilde{h_0}^{\rm odd}$.)  From those quantities he
constructs combinations which are invariant with respect to
diffeomorphisms on the hypersurfaces. Since the quantities, by
construction, are automatically independent of shift and lapse choice,
they are totally gauge invariant. Perturbation quantities which
decouple from others must have this property. Moncrief's method leads
to the same quantity $Q^{\rm odd}$ as that derived by Regge and 
Wheeler [if $Q^{\rm odd}$
is interpreted in an arbitrary gauge with (\ref{oddtrans2})].

In the case of even parity perturbations, the situation becomes much
more complicated.  Zerilli has derived a wave equation (the ``Zerilli
equation'') for a single decoupled quantity, by working with Fourier
transforms, that is, by assuming a time dependence $e^{-i\omega t}$.
We shall not be able to use Fourier transforms when dealing with the
nonlinear terms in second order perturbations, so we start by
re-deriving the Zerilli wave equation in the time domain. A point to
notice is that Zerilli introduces a function $R$ that is equivalent in
the time domain to $i\int H_1\,dt$. To avoid introducing an integral
we will be working with what amounts to the time derivative of
Zerilli's equations. Our time domain equations can be compared to the
frequency domain equations of Zerilli by replacing 
$\partial/\partial t$ 
in our expressions by $-i\omega$, and by replacing our 
${\ }^{(1)}\widehat{R}$ with Zerilli's 
$\widehat{R}_{LM}$, 
and our 
${\ }^{(1)}\chi$ 
by Zerilli's 
$-i\omega \widehat{K}_{LM}$.

From one of the vacuum Einstein equations we find that ${\ 
  }^{(1)}H_2^{\rm RW}={\ }^{(1)}H_0^{\rm RW}$. The remaining equations
can be broken into a first set that contains only first order
derivatives in $r$ and that can be written in the form
\begin{eqnarray}
\frac{\partial {\ }^{(1)}H_1^{\rm RW}} {\partial r} &= &
\frac{r}{r-2M}{\partial {\ }^{(1)}H_0^{\rm RW} \over \partial
t}+\frac{r}{r-2M}{\partial {\ }^{(1)}K^{\rm RW} \over \partial t}
-2\frac{M}{r(r-2M)} {\ }^{(1)}H_1^{\rm RW} \nonumber \\ \label{firstwt}
\frac{\partial^2 {\ }^{(1)}K^{\rm RW}} {\partial r \partial t} &= &
\frac{1}{r}{\partial {\ }^{(1)}H_0^{\rm RW} \over \partial
t}+\frac{\ell(\ell+1)}{2r^2} {\ }^{(1)}H_1^{\rm RW}-\frac{r-3M}{r
(r-2M)}{\partial {\ }^{(1)}K^{\rm RW} \over \partial t} \\
\frac{\partial^2 {\ }^{(1)}H_0^{\rm RW}} {\partial r \partial t} &= &
\frac{ r}{r-2M}{\partial^2 {\ }^{(1)}H_1^{\rm RW} \over \partial
t^2}+\frac{r-4M}{r(r-2M)}{\partial {\ }^{(1)}H_0^{\rm RW} \over
\partial
t}\nonumber\\
&&\hspace*{.4in}+\frac{3M-r}{r(r-2M)}{\partial {\ }^{(1)}K^{\rm RW} \over \partial
t}+\frac{\ell(\ell+1)}{2r^2} {\ }^{(1)}H_1^{\rm RW}
\nonumber\ .
\end{eqnarray}

We also obtain a second set of three equations containing second order
derivatives with respect to $r$. This makes a total of six equations
for the three unknown functions  ${\ }^{(1)}H^{\rm RW}$,
${\ }^{(1)}H_1^{\rm RW}$, and ${\ }^{(1)}K^{\rm RW}$, where
${\ }^{(1)}H^{\rm RW}\equiv{\ }^{(1)}H_2^{\rm RW}={\ }^{(1)}H_0^{\rm RW}$.
Compatibility of the system requires then that these equations are not
independent. One can show that by replacing the first three equations
in the second set, one obtains a single compatibility condition (named
the ``algebraic identity'' in Ref.\, \cite{zprl}). This can be written
as
\begin{eqnarray}
 \nonumber
& & -\frac{\ell(\ell+1)M}{r^2} {\ }^{(1)}H_1^{\rm RW} - 2
r\frac{\partial^2
 {\ }^{(1)}H_1^{\rm RW}}{\partial t^2}
 -\left[(\ell-1)(\ell+1)+\frac{6M}{r}\right]\frac{\partial{\ }^{(1)}
H^{\rm RW}}{\partial t} \nonumber
\\
 &+& \left[(\ell-1)(\ell+2)+\frac{2M(r-3M)}{r(r-2M)}
 \right]\frac{\partial{\ }^{(1)}K^{\rm RW}}{\partial t}  + \frac{2
 r^3}{r-2M}\frac{\partial^3 {\ }^{(1)}K^{\rm RW}}{\partial t^3} =0
\label{algid1}\ .
\end{eqnarray}
This equation together with (\ref{firstwt}), provides a set of four
differential equations for   ${\ }^{(1)}H^{\rm RW}$,
${\ }^{(1)}H_1^{\rm RW}$, and ${\ }^{(1)}K^{\rm RW}$. One can again
show that if (\ref{algid1}) and two of the equations (\ref{firstwt})
are satisfied, then the remaining equation is also satisfied.  We can
now take the $t$ derivative of the equations in
(\ref{firstwt})
and
use  (\ref{algid1}) to eliminate
${\ }^{(1)}H^{\rm RW},_ t$
in the second and third of the equations in 
(\ref{firstwt}). This reduces the system to two coupled linear partial
differential equations for 
${\ }^{(1)}H_1^{\rm RW}$
and 
${\ }^{(1)}K^{\rm RW}$, with $r$-dependent
coefficients, of first order in $r$ and second order in $t$. As shown
in Zerilli's paper this system can be ``diagonalized'' by the
transformation,
\begin{eqnarray}
\partial{\ }^{(1)}K^{\rm RW}/\partial t&=&f(r){\ }^{(1)}\chi +g(r)
{\ }^{(1)}\widehat{R} \label{Zer00}\\ {\ }^{(1)}H_1^{\rm RW}&=&
h(r){\ }^{(1)}\chi+k(r){\ }^{(1)}\widehat{R}\ ,
\label{Zer01}
\end{eqnarray}
where 
$$
f(r)=\frac{\lambda(\lambda+1)r^2 +3 \lambda M r + 
6 M^2 }{r^2(\lambda r-3M)} \;\;, \;\; g(r)=1
$$
\begin{equation}\label{handk}
h(r)= \frac{\lambda r^2-3 \lambda Mr-3M^2}{(r-2M)(\lambda r+3M)} 
\;\;,\;\;k(r)=
\frac{r^2}{r-2M} \;\;,\;\; \lambda = \frac{(\ell-1)(\ell+2)}{2}.
\end{equation}
A necessary condition for the Einstein equations to be satisfied is 
that
${\ }^{(1)}\chi$ and 
 ${\ }^{(1)}\widehat{R}$ 
satisfy
\begin{equation}\label{zersys}
\frac{\partial {\ }^{(1)}\chi} {\partial r^*} = {\ }^{(1)}\widehat{R}
\;\;,\;\; \frac{\partial {\ }^{(1)}\widehat{R}} {\partial r^*} =
\left[V(r^*) + \frac{\partial^2 } {\partial t^2} \right] {\ }^{(1)}\chi
\end{equation}
where
\begin{equation}
V(r^*) = 2 \left(1 -{2 M \over r}\right) { \lambda^2 r^2
\left[(\lambda+1) r + 3 M \right] + 9 M^2 (\lambda r +M) \over r^3
(\lambda r + 3 M)^2 }\ ,
\end{equation}
and, as in (\ref{rweq}), 
\begin{equation}
r^*=r+2M \ln[r/(2M)-1]\ .
\end{equation}
Equations (\ref{zersys}) imply that ${\ }^{(1)}\chi$ satisfies
the ``Zerilli equation''
\begin{equation}
{\partial^2 {\ }^{(1)}\chi \over \partial {r^*}^2} - {\partial^2
{\ }^{(1)}\chi \over \partial t^2} -V(r^*) {\ }^{(1)}\chi =0\ .
\label{Zer21}
\end{equation}

A necessary condition for perturbations 
${\ }^{(1)}H_1^{RW}$ and
${\ }^{(1)}K^{RW}$ 
to satisfy Einstein's equations, is that they satisfy the relations
in the system of equations
(\ref{Zer00})--(\ref{zersys}), for 
${\ }^{(1)}\chi$ satisfying 
(\ref{Zer21}). But in arriving at this system we have differentiated
Einstein's equations with respect to $t$. It will be useful in what
follows to set down here two of the original (not time differentiated)
Einstein equations:
\begin{equation}\label{Hdoteq}
r\frac{\partial {\ }^{(1)}H_1^{RW}  }{\partial t}
=\frac{2M}{r}{\ }^{(1)}H^{RW}
+\left(r-2M\right)
\left(
\frac{\partial {\ }^{(1)}H^{RW}  }{\partial r}
-\frac{\partial {\ }^{(1)}K^{RW}  }{\partial r}
\right)
\end{equation}
\begin{displaymath}
\frac{r^3}{r-2M}\frac{\partial^2 {\ }^{(1)}K^{RW}  }{\partial t^2}
=
r(r-2M)\frac{\partial^2 {\ }^{(1)}K^{RW}  }{\partial r^2}
+
2(2r-3M)\frac{\partial {\ }^{(1)}K^{RW}  }{\partial r}
\end{displaymath}\begin{equation}
\label{Kdotdoteq}
-
2(r-2M)\frac{\partial {\ }^{(1)}H^{RW}  }{\partial r}
+
2r\frac{\partial {\ }^{(1)}H_1^{RW}  }{\partial t}
+
(\ell-1)(\ell+2){\ }^{(1)}K^{RW}
-2{\ }^{(1)}H^{RW}\ .
\end{equation}

%%%%%%%%%%%%%%%%%%%%%%%%%%%
%xxxxxxxxxxxxxxxxxxx
%%%%%%%%%%%%%%%%%%%%%%%%%%

Once (\ref{Zer21}) is solved, the values of all the RW perturbations,
can be found.  One finds ${\ }^{(1)}\widehat{R}$ from the first of the
relations in (\ref{zersys}), and then from (\ref{Zer00}) and
(\ref{Zer01}) one finds ${\ }^{(1)}H_1^{\rm RW}$ and ${\ 
  }^{(1)}\dot{K}^{\rm RW}$.  The solution for ${\ }^{(1)}\dot{H}^{\rm
  RW}$ is then found from the second of the equations in
(\ref{firstwt}). Lastly the ``integration constants'' in finding ${\ 
  }^{(1)}{K}^{\rm RW},{\ }^{(1)}{H}^{\rm RW}$ from ${\ 
  }^{(1)}\dot{K}^{\rm RW},{\ }^{(1)}\dot{H}^{\rm RW}$ are known from
the metric on the $t=0$ hypersurface, which fixes the initial values
of ${\ }^{(1)}{K}^{\rm RW},{\ }^{(1)}{H}^{\rm RW}$.

The Zerilli equation requires Cauchy data, values of ${\ }^{(1)}\chi$ 
and of $\partial{\ }^{(1)}\chi/\partial t$, at
$t=0$, and the Cauchy data must originate in an initial value solution
of Einstein's equations, i.e., our solution, in
some arbitrary gauge, for the 3-metric $\gamma_{ij}$  and the extrinsic 
curvature $K_{ij}$. 
From these 
we immediately get the $t=0$
values of 
${\ }^{(1)}\widetilde{H}_2,
{\ }^{(1)}\widetilde{h}_1,
{\ }^{(1)}\widetilde{K}$
and
${\ }^{(1)}\widetilde{G}$.
To proceed further it is convenient, though not necessary, to choose to
view our initial data to be in a gauge with 
${\ }^{(1)}\widetilde{h}_0={\ }^{(1)}\widetilde{H}_0=
{\ }^{(1)}\widetilde{H}_1=0$.
In this gauge
(\ref{Kgamdot})
gives us the time derivatives of
${\ }^{(1)}\widetilde{H}_2,
{\ }^{(1)}\widetilde{h}_1,
{\ }^{(1)}\widetilde{K},
{\ }^{(1)}\widetilde{G}$.
immediately from 
$K_{ij}$. Using this information about the initial 3-geometry
perturbations, and
the definitions 
(\ref{eventrans1})--(\ref{eventrans3}), we can find
the initial values of 
${\ }^{(1)}H^{\rm RW}$,
${\ }^{(1)}H_1^{\rm RW}$,
${\ }^{(1)}K^{\rm RW}$
and of the time derivatives
${\ }^{(1)}\dot{H}^{\rm RW}$,
${\ }^{(1)}\dot{K}^{\rm RW}$.
As an example, from the Misner initial value solution, which 
has vanishing initial extrinsic curvature, if we choose 
${\ }^{(1)}\widetilde{h}_0={\ }^{(1)}\widetilde{H}_0=
{\ }^{(1)}\widetilde{H}_1=0$, we 
have
\begin{displaymath}
{\ }^{(1)}\widetilde{H}_2={\ }^{(1)}\widetilde{K}=
2\kappa_2\left(
1-\frac{2M}{R}
\right)^{-1} \left(
\frac{2M}{R}
\right)^3\sqrt{\frac{4\pi}{5}}
\end{displaymath}\begin{equation}\label{miscauchy}
{\ }^{(1)}\widetilde{G}={\ }^{(1)}\widetilde{h}_1=
{\ }^{(1)}\widetilde{H}_2,_t={\ }^{(1)}\widetilde{K},_t=
{\ }^{(1)}\widetilde{G},_t={\ }^{(1)}\widetilde{h}_1,_t=0
\ .
\end{equation}
With these we can solve (\ref{Zer00})  and (\ref{Zer01}) 
for ${\ }^{(1)}\chi$.
It remains to find ${\ }^{(1)}\chi,_t$, 
which requires
${\ }^{(1)}H_1^{\rm RW},_t$ and 
${\ }^{(1)}K^{\rm RW},_{tt}$.
With the values of the RW perturbations, and derivatives, already found,
${\ }^{(1)}H_1^{\rm RW},_t$ 
follows from (\ref{Hdoteq}),
and 
${\ }^{(1)}K^{\rm RW},_{tt}$
from 
(\ref{Kdotdoteq}).

The above development of first order perturbation theory gives the
time-domain Zerilli formalism in terms of the field variable 
${\ }^{(1)}\chi$, defined [see (\ref{Zer00}) --- (\ref{zersys})] by
\begin{equation}
\label{chideKH1}
{\ }^{(1)}\chi= {r-2M \over \lambda r +3M} \left[
{r^2 \over r-2M}{\partial {\ }^{(1)}{K}^{\rm RW} \over \partial
t}-{\ }^{(1)}H_1^{\rm RW}
\right]\ .
\end{equation}
This development is important here because it will be used as a
pattern for the second order formalism. For the first order
computations themselves, however, it is convenient not actually to use
${\ }^{(1)}\chi$, but rather to use a field variable 
${\ }^{(1)}\psi$ 
that is
roughly
equivalent to the time integral of ${\ }^{(1)}\chi$.

To introduce ${\ }^{(1)}\psi$ we start by using the second
relationship in (\ref{firstwt}) to replace ${\ }^{(1)}H_1^{\rm RW}$ in
(\ref{chideKH1}), with the result
\begin{equation}\label{Mondef2}
 {\ }^{(1)}\chi=
\frac{2r(r-2M)}{\ell(\ell+1) (\lambda r +3M)}\left[
{\ }^{(1)}\dot{H}^{\rm RW}
-r\frac{\partial {\ }^{(1)}\dot{K}^{\rm RW}}{\partial r}
-\frac{r-3M}{r-2M}{\ }^{(1)}\dot{K}^{\rm RW}
\right]
+\frac{r^2}{\lambda r +3 M}{\ }^{(1)}\dot{K}^{\rm RW}\ .
\end{equation}
This leads us to define
\begin{equation}\label{psi1def}
 {\ }^{(1)}\psi\equiv
\frac{2r(r-2M)}{\ell(\ell+1) (\lambda r +3M)}\left[
{\ }^{(1)}H^{\rm RW}
-r\frac{\partial {\ }^{(1)}K^{\rm RW}}{\partial r}
-\frac{r-3M}{r-2M}{\ }^{(1)}K^{\rm RW}
\right]
+\frac{r^2}{\lambda r +3 M}{\ }^{(1)}K^{\rm RW}\ ,
\end{equation}
which, from (\ref{Mondef2}) satisfies
\begin{equation}
   {\ }^{(1)}\dot{\psi}={\ }^{(1)}\chi\ .
\end{equation}
It is straightforward to show that 
   ${\ }^{(1)}{\psi}$, like ${\ }^{(1)}\chi$,
satisfies the Zerilli equation 
 \begin{equation}
{\partial^2 {\ }^{(1)}\psi \over \partial {r^*}^2} - {\partial^2
{\ }^{(1)}\psi \over \partial t^2} -V(r^*) {\ }^{(1)}\psi =0\ .
\label{Zer21a}
\end{equation}
It will be useful for the analysis below of asymptotic behavior to note that
in terms of ${\ }^{(1)}{\psi}$ the equivalents of 
(\ref{Zer00}),
(\ref{Zer01}), and
(\ref{zersys})
can be written
\begin{eqnarray}
\label{Kdepsi}
  {\ }^{(1)}{K}^{RW}   
& = & f(r) {\ }^{(1)}\psi 
+ \left( 1 - {2 M \over r} \right) 
{\partial {\ }^{(1)} \psi \over \partial r}   \\
\label{H1depsi}
 {\ }^{(1)}{H}_1^{RW} & = & h(r)  {\partial {\ }^{(1)} \psi
 \over \partial t} + r {\partial^2 {\ }^{(1)} \psi \over \partial t
 \partial r} \\
\label{Hdepsi}
 {\ }^{(1)}{H}^{RW}  
 &= & {\partial {\; } \over \partial r}
\left[ \left(1 - {2M \over r}\right) 
h(r)  {\ }^{(1)} \psi   + r {\partial {\ }^{(1)} \psi \over \partial r}
\right]  
-  {\ }^{(1)}{K}^{RW}\ .
\end{eqnarray}

The wavefunction ${\ }^{(1)}{\psi}$ is very closely related to the
variable used by Moncrief \cite{moncrief}, which is defined in terms of
perturbations in an arbitrary gauge to be
\begin{displaymath}
{\ }^{(1)}\psi_{\rm Monc}\equiv
\frac{2r(r-2M)}{\ell(\ell+1)(\lambda r +3M)}\left[
{\ }^{(1)}\widetilde{H}_2
-r\frac{\partial {\ }^{(1)}\widetilde{K}}{\partial r}
-\frac{r-3M}{r-2M}{\ }^{(1)}\widetilde{K}
\right]
\end{displaymath}
\begin{equation}\label{Mondef}
+{r^2 \over \lambda r +3M} \left[{\ }^{(1)}\widetilde{K}
+(r-2M)
\left(
\frac{\partial {\ }^{(1)}\widetilde{G}}{\partial r}
-{2 \over r^2}{\ }^{(1)}\widetilde{h}_1
\right) \right] .
\end{equation}
(Note that our expression for ${\ }^{(1)}\psi_{\rm Monc}$ must be
multiplied by $\ell(\ell+1)$ to get the expression in Moncrief's
paper.)  It is clear that ${\ }^{(1)}\psi_{\rm Monc}$ reduces to ${\ 
  }^{(1)}\psi$, when one specializes to the RW gauge and with the use
of one of the vacuum field equations sets ${\ }^{(1)}{H}_2^{\rm
  RW}={\ }^{(1)}{H}^{\rm RW}$.  The great advantage of Moncrief's
formulation in (\ref{Mondef}) is that ${\ }^{(1)}\psi_{\rm Monc}$ is
defined entirely in terms of components of the 3-geometry. This
greatly simplifies finding Cauchy data from an initial value solution.

The simple relationship of the Zerilli and the Moncrief wavefunctions
disappears in the presence of real source terms, or of the effective
source terms that appear in the second order equations.  In our second
order development we could have followed either the pattern leading to
a Zerilli equation for a second order variable analogous to ${\ 
  }^{(1)}\chi$, or we could have followed the pattern of Moncrief's
derivation \cite{moncrief} to find a Zerilli equation for a variable
analogous to ${\ }^{(1)}\psi_{\rm Monc}$. The former choice has the
disadvantage that it is related to the metric perturbations by an
additional time derivative, but it has the advantage of 
greater simplicity than the Moncrief approach. The complexity
of the second order equations makes simplicity the more important 
consideration, and we have chosen to work with a second order equivalent
of ${\ }^{(1)}\chi$.

\subsection{Radiation}
 
The solution for first order perturbations does not directly give us
the flux of radiated power. To analyze the radiation we must go to a
coordinate system in which the metric has a manifestly asymptotically
flat (AF) form, in which the deviations from the Schwarzschild metric
fall off with radius as follows.(See Ref.\, \cite{mtw}, Chap.\,19.)
\begin{displaymath}
\delta g_{00}, \delta g_{01}, \delta g_{11}, ={\cal O}(r^{-2})\;\;\;
\delta g_{02},\delta g_{03},\delta g_{12},\delta g_{13}={\cal
O}(r^{-1})\;\;\;
\end{displaymath}\begin{equation}\label{AFrules}
\delta g_{22},\delta g_{23},\delta g_{33}={\cal O}(r)\ .
\end{equation}
In this coordinate system the information about gravitational
radiation is carried by the transverse metric components $\delta
g_{22},\delta g_{23},\delta g_{33}$, and these perturbations
will have the form $r\times$function of $t-r^*$.

These are requirements on the $r$ dependence of the metric functions
for the coordinate system to be AF. If it is to be AF to first order
then these conditions must be satisfied by the metric perturbations to
first order.  The RW gauge is not AF. There are two rather different
ways in which we can extract radiation information from our RW
results. The first method is 
to use the gauge invariant expressions for the first order 
${\ }^{(1)}\chi$
and
$Q^{\rm odd}$,
and to look at the form they take in the radiation zone, for 
an asymptotically flat gauge.
To first order,
the AF gauge conditions requires that AF perturbations fall of as 
\begin{equation}
H_0^{\rm AF},H_1^{\rm AF},H_2^{\rm AF}
\sim r^{-2}\;\;\;
h_0^{\rm AF}, h_1^{\rm AF},
G^{\rm AF},K^{\rm AF} 
\sim r^{-1}
\end{equation}
\begin{equation}
h_0^{\rm AF,odd}, h_1^{\rm AF,odd}\sim r^{-1}\;\;\;
h_1^{\rm AF,odd}\sim r
\end{equation}  
For first order odd parity perturbations, 
(\ref{oddtrans1}),
(\ref{RWQ})
in the radiation zone, then give us,
\begin{equation}\label{oddarbitrary}
Q^{\rm odd}\approx r^{-1}{\ }^{(1)}h_1^{\rm RW, odd}
\approx 
-\textstyle{\frac{1}{2}}r^{-1}{\ }^{(1)}\dot{h}_2^{\rm AF,odd}\ ,
\end{equation}
where the dot means differentiation with respect to time, or to retarded 
time.
Of particular interest is the power radiated. From the asymptotic
form of the perturbations, one finds \cite{cpm}
\begin{equation}
{\rm Power}= \epsilon^2\frac{1}{16\pi}\,\frac{(\ell+2)!}{(\ell-2)!}
\left(Q^{\rm odd}\right)^2\ .
\end{equation}
For even parity, in the radiation zone, using (\ref{chideKH1}), the
relation equivalent to
(\ref{oddarbitrary}) 
is:
\begin{displaymath}
{\ }^{(1)}\chi 
\approx\textstyle{\frac{1}{\lambda}}\left(
r{\ }^{(1)}\dot{K}^{\rm RW}-{\ }^{(1)}H_1^{\rm RW}
\right) \;,
\end{displaymath}
or, using the first order gauge transformation equations,
\begin{equation}\label{chiAF}
{\ }^{(1)}\chi \approx\frac{r}{\lambda}
\left(
{\ }^{(1)}\dot{K}^{\rm AF}
-{\ }^{(1)}\dot{G}^{\rm AF}
\right)\ ,
\end{equation}
and from this asymptotic form, one finds that the radiated power
is 
\begin{equation}
{\rm Power}=\epsilon^2\frac{\pi}{4}\,\frac{(\ell+2)!}{(\ell-2)!}\,
\frac{1}{(2\ell+1)^2}\,\chi^2\ .
\end{equation}

A very different procedure for extracting radiation is to perform an
explicit gauge transformation from the RW gauge to a gauge that is AF.
Since this will be the basis of the approach we use for the second
order problem, we illustrate it here.  For simplicity of presentation,
we limit attention to the even-parity $\ell=2$ case, the case of
interest for second order calculations. We shall also write the
perturbative functions in terms of ${\ }^{(1)}\psi(t,r)$ instead of
${\ }^{(1)}\chi(t,r)$ to avoid having to deal with $ {\ }^{(1)}K^{\rm
RW},_t$, and $ {\ }^{(1)}H^{\rm RW},_t$, instead of $ {\ }^{(1)}K^{\rm
RW}$, and $ {\ }^{(1)}H^{\rm RW}$
We start by noticing that the solution to the Zerilli equation
(\ref{Zer21a}), for outgoing radiation, can be written in the form of
an expansion in powers of $r$ in which the coefficients are functions
of retarded time $t-r^*$. 
That is, we write ${\ }^{(1)}\psi(t,r)$
in the form 
\begin{equation}\label{F1exp1}
{\ }^{(1)}\psi(t,r) = {\ }^{(1)}F_a(t-r^*) + {\ }^{(1)}F_b(t-r^*)/r 
+{\ }^{(1)}F_c(t-r^*)/r^2 + {\cal O}(1/r^3)\ .
\end{equation}
Substituting this {\it ansatz} in (\ref{Zer21a}) one can show
that ${\ }^{(1)}F_b$, ${\ }^{(1)}F_c, \cdots$ are determined once, e.g., ${\
}^{(1)}F_a$ is fixed. For the first terms in the expansion
(\ref{F1exp1}), these relations may be conveniently expressed in terms
of a suitably defined function $F(t)$
as,
\begin{equation}\label{Fexpan2}
{\ }^{(1)}F_a(t)= {1 \over \lambda+1}  {\ }^{(1)}F''(t) 
 \;\;\;,\;\;\;
{\ }^{(1)}F_b(t)=  {\ }^{(1)}F'(t) 
 \;\;\;,\;\;\;
 {\ }^{(1)}F_c(t)= {\lambda \over 2} {\ }^{(1)}F(t) - {3 M (\lambda+2)
 \over 2 \lambda(\lambda+1)}  {\ }^{(1)}F'(t)\ ,
 \end{equation} 
where $F'(x)=dF(x)/dx$, $F''(x)=d^2F(x)/dx^2$, etc. In particular, for
the quadrupole contribution $\ell=2$, we have,
\begin{equation}\label{Fexpan}
{\ }^{(1)}\psi
=\frac{1}{3}{\ }^{(1)}F''(t-r^*)+\frac{1}{r}{\ }^{(1)}F'(t-r^*)
+\frac{1}{r^2}\left[
{\ }^{(1)}F-M{\ }^{(1)}F'
\right]+\cdots\ .
\end{equation}  
The form of 
${\ }^{(1)}\psi$ and hence of 
${\ }^{(1)}F(t)$ will, of course, be determined by the Cauchy data used
in the solution of (\ref{Zer21a}).

For $r\gg M$ we may obtain the asymptotic RW perturbation functions in
terms of ${\ }^{(1)}\psi$, by replacing in (\ref{Kdepsi}),
(\ref{H1depsi}), (\ref{Hdepsi}) the asymptotic form of
${\ }^{(1)}\psi$. Restricting again for simplicity to $\ell=2$, we
find:
\begin{eqnarray}
{K}^{\rm RW} &= & {\partial {\ }^{(1)}\psi \over \partial r} + {3 \over r}
{\ }^{(1)}\psi - {2M \over r} {\partial {\ }^{(1)}\psi \over \partial r} +
O(1/r^2) \nonumber \\
{H}_0^{\rm RW} &= & r {\partial^2 {\ }^{(1)}\psi \over \partial r^2}-2M
{\partial^2 {\ }^{(1)}\psi \over \partial r^2} + {\partial
{\ }^{(1)}\psi \over \partial r} - {3 \over r} {\ }^{(1)}\psi - {5M \over 2r}
{\partial {\ }^{(1)}\psi \over \partial r} + O(1/r^2) \nonumber \\
H_1^{\rm RW} &= & r {\partial^2 {\ }^{(1)}\psi \over \partial r \partial t} +
{\partial {\ }^{(1)}\psi \over \partial t}  - {5M \over 2r} {\partial
{\ }^{(1)}\psi \over \partial t} + O(1/r^2)  \\
H_2^{\rm RW} &= & H_0^{\rm RW} \nonumber\ .
\end{eqnarray}

From these relations, and from (\ref{Fexpan}), we see that the RW
perturbations diverge for $r\rightarrow\infty$.  We can view these
divergent perturbations as an indication that the RW gauge is not
asymptotically flat.  (Here we are considering the RW quantities as
perturbations in a specific gauge, the RW gauge; we usually view them
as gauge invariant combinations of perturbations). We can explicitly
perform a gauge transformation, analogous to that in (\ref{gauge2p}),
to take the perturbations from RW gauge to AF gauge:
\begin{eqnarray}
{\ }^{(1)}H_0^{\rm AF} & = &{\ }^{(1)}H_0^{\rm RW}  + {2M \over
r(r-2M)} {\ }^{(1)}\alpha_1 +2 {\partial {\ }^{(1)}\alpha_0
\over \partial t}  \nonumber \\
{\ }^{(1)}H_1^{\rm AF} & = &{\ }^{(1)}H_1^{\rm RW}  - {r \over r-2M}
{\partial {\ }^{(1)}\alpha_1\over \partial
t}  +{r-2M \over r} {\partial {\ }^{(1)}\alpha_0\over \partial r}
\nonumber \\ {\ }^{(1)}h^{\rm AF}_0 & = &  - r^2 {\partial
{\ }^{(1)}\alpha_2\over \partial t} +{r-2M \over r} {\ }^{(1)}\alpha_0
\nonumber \\ \label{gauge3}
{\ }^{(1)}H^{\rm AF}_2 & = & {\ }^{(1)}H_2^{\rm RW}+ {2M \over r(r-2M)}
{\ }^{(1)}\alpha_1 -2 {\partial {\ }^{(1)}\alpha_1\over \partial r }
\\
{\ }^{(1)}h^{\rm AF}_1 & = & - {r \over r-2M} {\ }^{(1)}\alpha_1 - r^2 {\partial {\ }^{(1)}\alpha_2\over
\partial r}  \nonumber \\
{\ }^{(1)}G^{\rm AF} & = &  - 2 {\ }^{(1)}\alpha_2  \nonumber \\
{\ }^{(1)}K^{\rm AF} & = & {\ }^{(1)}K^{\rm RW} 
- {2 \over r} {\ }^{(1)}\alpha_1  \nonumber \ ,
\end{eqnarray}
where we are denoting with 
${\ }^{(1)}\alpha_0$,
${\ }^{(1)}\alpha_1$,
${\ }^{(1)}\alpha_2$,
the specification of gauge vector ${\ }^{(1)}\xi$, the role 
played by
${\ }^{(1)}A_0$,
${\ }^{(1)}A_1$,
${\ }^{(1)}A_2$,
in (\ref{gauge2p}).
To cancel the leading term in ${\ }^{(1)}K^{\rm RW}$, and force the
appropriate asymptotic behavior for $K^{\rm AF}$ we choose
\begin{equation}\label{alpha1}
{\ }^{(1)}{\alpha}_1 = {r \over 2}{\partial {\ }^{(1)}\psi \over \partial r}\ .
\end{equation}
Similarly, the leading divergences in 
${\ }^{(1)}H_0^{\rm RW}$
and
${\ }^{(1)}H_2^{\rm RW}$
are cancelled 
by the choice
\begin{equation}
{\ }^{(1)}\alpha_0 = -{r \over 2}{\partial {\ }^{(1)}\psi \over \partial t}
\end{equation}
Finally, to avoid a ``bad'' behavior in ${\ }^{(1)}h_0$ 
and ${\ }^{(1)}h_1$ we choose,
\begin{equation}
{\ }^{(1)}{\alpha}_2 = -{1 \over 2 r} {\ }^{(1)}\psi\ .
\end{equation}

This procedure may be iterated to the desired degree of accuracy.
Since we are only interested in showing that there is a choice of gauge
functions that carries the RW gauge to an asymptotically flat gauge,
the computations are simplified if we consider from the start the
expression for 
the RW gauge perturbations in terms of ${\ }^{(1)}F$.  To put the
asymptotic form of the full metric in an ``asymptotically flat'' gauge,
we need an expansion of 
${\ }^{(1)}\chi$
up to terms of order $1/r^3$. The
corresponding asymptotic forms for 
${\ }^{(1)}H^{\rm RW}$,
${\ }^{(1)}H_1^{\rm RW}$,
and
${\ }^{(1)}K^{\rm RW}$
are then,
\begin{eqnarray}
{\ }^{(1)}{H}^{\rm RW} & = & {r\over 3} {\partial^4 {\ }^{(1)}F \over
\partial t^4}(t-r^*)
        + {2 M\over 3} {\partial^4 {\ }^{(1)}F \over \partial t^4}(t-r^*)
        +{2\over 3} {\partial^3 {\ }^{(1)}F \over \partial t^3}(t-r^*)
 \nonumber \\
& & + {1 \over r}  \left[ {4 M^2\over 3} 
         {\partial^4 {\ }^{(1)}F \over \partial t^4}(t-r^*)
          + {11 M\over 6} {\partial^3 {\ }^{(1)}F \over \partial t^3}(t-r^*)
        +{\partial^2 {\ }^{(1)}F \over \partial t^2}(t-r^*) \right]
          +\mbox{O}(1/r^2) \nonumber \\
{\ }^{(1)}H_1^{\rm RW}& = & -{r\over 3} {\partial^4 {\ }^{(1)}F \over
\partial t^4}(t-r^*)
        - {2 M\over 3} {\partial^4 {\ }^{(1)}F \over \partial t^4}(t-r^*)
        -{2\over 3} {\partial^3 {\ }^{(1)}F \over \partial t^3}(t-r^*) 
\label{RWofF}\\
& & -{1 \over r}  \left[ {4 M^2\over 3} 
         {\partial^4 {\ }^{(1)}F \over \partial t^4}(t-r^*)
          + {11 M\over 6} {\partial^3 {\ }^{(1)}F \over \partial t^3}(t-r^*)
        +{\partial^2 {\ }^{(1)}F \over \partial t^2}(t-r^*) \right]
         +\mbox{O}(1/r^2) \nonumber \\
{\ }^{(1)}{K}^{\rm RW} & = & -{1\over 3} {\partial^3 {\ }^{(1)}F \over
\partial t^3}(t-r^*)
         +{1 \over r^2}  \left[   
         {\partial {\ }^{(1)}F \over \partial t}(t-r^*)
        + {  M\over 2} {\partial^2 {\ }^{(1)}F \over \partial
        t^2}(t-r^*) \right]
        \nonumber \\
& & +{1 \over r^3}  \left[   
           {\ }^{(1)}F  (t-r^*)
        + {  M\over 2} {\partial {\ }^{(1)}F \over \partial t}(t-r^*) \right]
      +\mbox{O}(1/r^2) \nonumber\ .
\end{eqnarray}

By using these expansions on the right hand side of (\ref{gauge3}) we
then find for ${\ }^{(1)}\alpha_0$, ${\ }^{(1)}\alpha_1$ 
and ${\ }^{(1)}\alpha_2$ the asymptotic
expansions,
%%%%%%%%%%%%%%%%%%%%%%%%%%%%%%%%%
\begin{eqnarray}
{\ }^{(1)}\alpha_0 & = & -{r \over 6}{\partial^3 {\ }^{(1)}F \over
\partial t^3}(t-r^*)
   -{M \over 3}{\partial^3 {\ }^{(1)}F \over \partial t^3}(t-r^*)
-{1\over 3}{\partial^2 {\ }^{(1)}F \over \partial t^2}(t-r^*) \nonumber \\
%%%%%%
& & + {1 \over r}  \left[   
         -{1 \over 2}{\partial {\ }^{(1)}F \over \partial t}(t-r^*)
        - { 3 M\over 4} {\partial^2 {\ }^{(1)}F \over \partial t^2}(t-r^*) 
- { 2 M^2\over 3} {\partial^3 {\ }^{(1)}F \over \partial t^3}(t-r^*)\right]
 \nonumber \\
%%%%%%%%%%%%%%%%
& & + {1 \over r^2}  \left[ -{1 \over 2} {\ }^{(1)}F(t-r^*)   
         -{3M \over 4}{\partial {\ }^{(1)}F \over \partial t}(t-r^*)
        - { 3 M^2\over 2} {\partial^2 {\ }^{(1)}F \over \partial
        t^2}(t-r^*)\right.
\nonumber\\ 
%%%%%%%%%%%%%%%%%%%%%
& &\left.- { 4 M^3\over 3} {\partial^3 {\ }^{(1)}F \over \partial
t^3}(t-r^*)\right] + {1 \over r^3} {\cal A}_0(t-r^*) + \mbox{O}(1/r^4)
\label{avsF}\\
%%%%%%%%%%%%%%%%%%%%
{\ }^{(1)}{\alpha}_1 & = &  -{r \over 6}{\partial^3 {\ }^{(1)}F \over
\partial t^3}(t-r^*)
   -{1 \over 2}{\partial^2 {\ }^{(1)}F \over \partial t^2}(t-r^*)
 + {1 \over r}  \left[   
         -{1 \over 2}{\partial {\ }^{(1)}F \over \partial t}(t-r^*)
        + {  M\over 4} {\partial^2 {\ }^{(1)}F \over \partial t^2}(t-r^*) \right]
 \nonumber \\
%%%%%%%%%%%%%%%%%%%%
 & & + {1 \over r^3} {\cal A}_1(t-r^*) + \mbox{O}(1/r^4) \nonumber\\
%%%%%%%%%%%%%%%%%%%%%%%
{\ }^{(1)}{\alpha}_2 & = & -{1 \over 6 r} 
{\partial^2 {\ }^{(1)}F \over \partial t^2}(t-r^*)
    -{1 \over 3 r^2} {\partial {\ }^{(1)}F \over \partial t}(t-r^*) 
       +{1 \over r^3}{\cal A}_2(t-r^*) + +\mbox{O}(1/r^4) \nonumber 
\end{eqnarray}

With these expansions, and 
(\ref{RWofF}) in (\ref{gauge3}), we finally find
\begin{eqnarray}
{\ }^{(1)}{H}^{\rm AF}_0 & = &  {1 \over r^3} \left[{16 \over 3} M^4
{\ }^{(1)}F''''(t-r^*)
    + {69 \over 8} M^3 {\ }^{(1)}F'''(t-r^*) + {5 \over 4} M^2
    {\ }^{(1)}F''(t-r^*) \right. \nonumber \\
& & \left.
     +{\ }^{(1)}F(t-r^*) +2{{\cal A}_0}'(t-r^*) + {\cal{F}}''(t-r^*)
     \right] +\mbox{O}(1/r^4) \nonumber \\
{\ }^{(1)}H^{\rm AF}_1 & = &  {1 \over r^3} \left[-{8 \over 3} M^4
{\ }^{(1)}F''''(t-r^*)
    - {45 \over 8} M^3 {\ }^{(1)}F'''(t-r^*) + {3 \over 4} M^2
    {\ }^{(1)}F''(t-r^*) \right. \nonumber \\
& & \left.
     +{\ }^{(1)}F(t-r^*) - {{\cal A}_0}'(t-r^*)- {{\cal A}_1}'(t-r^*) -
     {\cal{F}}''(t-r^*) \right] +\mbox{O}(1/r^4) \nonumber \\
{\ }^{(1)}{H}^{\rm AF}_2 & = & {1 \over r^3} \left[
    - {21 \over 8} M^3 {\ }^{(1)}F'''(t-r^*) - {11 \over 4} M^2
    {\ }^{(1)}F''(t-r^*) \right. \label{AFofF}\\
& & \left.
     +{\ }^{(1)}F(t-r^*) +2{\cal A}_1'(t-r^*) + {\cal{F}}''(t-r^*)
     \right] +\mbox{O}(1/r^4) \nonumber \\
{\ }^{(1)}h^{\rm AF}_0 & = & {1 \over r} \left[
    - {M \over 12} {\ }^{(1)}F'''(t-r^*) - {1 \over 2} {\ }^{(1)}F'(t-r^*) 
        -{\cal A}'_2(t-r^*) \right]   +\mbox{O}(1/r^2) \nonumber \\
 \label{gauge4}
{\ }^{(1)}{h}^{\rm AF}_1 & = &   {1 \over r} \left[
     {M \over 12} {\ }^{(1)}F'''(t-r^*) - {1 \over 6} {\ }^{(1)}F'(t-r^*) 
        +{\cal A}'_2(t-r^*) \right] +\mbox{O}(1/r^2) \nonumber \\
 {\ }^{(1)}{G}^{\rm AF} & = &  {1 \over 3r}{\ }^{(1)}F''(t-r^*)+ {2
 \over 3r^2}{\ }^{(1)}F'(t-r^*) - {2 \over r^3}{\cal A}_2(t-r^*)
 +\mbox{O}(1/r^4) \nonumber \\
 {\ }^{(1)}{K}^{\rm AF} & = &  {1 \over r}{\ }^{(1)}F''(t-r^*)+ {2
 \over r^2}{\ }^{(1)}F'(t-r^*) + {1\over 2r^3}\left[M
 {\ }^{(1)}F'(t-r^*) + 2{\ }^{(1)}F(t-r^*) \right]    +\mbox{O}(1/r^4)
 \nonumber
  \end{eqnarray}

In all these expressions $\cal{F}$ is the contribution of order
$1/r^4$ in ${\ }^{(1)}\chi$, and we notice that ${\cal A}_0$,${\cal A}_1$
and ${\cal A}_2$ are not determined by the condition of asymptotic
flatness, and can be freely specified.

For computational purposes, it is useful to have expressions for
the asymptotic forms for 
${\ }^{(1)}\alpha_0$,${\ }^{(1)}\alpha_1$
and ${\ }^{(1)}\alpha_2$
in terms of
${\ }^{(1)}\psi$, instead of ${\ }^{(1)}F$ as in (\ref{AFofF}). This inversion is
rather cumbersome, because a given term in 
${\ }^{(1)}\psi$
contributes to
many orders. The leading behavior found in (\ref{AFofF}) is
reproduced by, 
\begin{eqnarray} 
{\ }^{(1)}{\alpha}_2 & = & -{1 \over 2 r} {\ }^{(1)}\psi+ {1\over 2r^2}
\int {\ }^{(1)}\psi dt \nonumber \\ 
{\ }^{(1)}\alpha_0 & = & -{r \over 2}{\partial
{\ }^{(1)}\psi \over \partial t} + {1 \over 2}{\ }^{(1)}\psi - M {\partial
{\ }^{(1)}\psi \over \partial t} \label{firstgavec} \\ 
{\ }^{(1)}{\alpha}_1 & = & {r \over
2}{\partial {\ }^{(1)}\psi \over \partial r} 
+M {\partial {\ }^{(1)}\psi \over
\partial t} \nonumber 
\end{eqnarray} 
but more terms need to be specified to
achieve the same order of accuracy as in (\ref{AFofF}). In practice one
needs only the leading terms in (\ref{AFofF}), and having solved
for ${\ }^{(1)}\psi$, one can find ${\ }^{(1)}F$, to needed accuracy,
from the ${\cal O}(r^0)$
and
${\cal O}(r^{-1})$
parts of ${\ }^{(1)}\psi$.

We have shown that there are gauge transformations such that the
metric functions can be put in the asymptotic form, and in which 
the gravitational wave amplitudes (the ${\cal O}(r^0)$ terms in 
${\ }^{(1)}F^{\rm AF}$ and ${\ }^{(1)}G^{\rm AF}$) can be found 
from  ${\ }^{(1)}\psi$\ , or, up to an irrelevant, $r$ dependent
`integration constant', in terms of ${\ }^{(1)}\chi$
In the next section we  explore the possibility of obtaining similar
results in the second order perturbations.

\section{Second order perturbations}

\subsection{Wave Equations}
We now turn to the Einstein equations at second order in the perturbations,
and find the first order formalism of the preceding section can
be modified to handle higher orders.
In the second order Einstein equations (\ref{second}),
\begin{equation}\label{secordeq}
L_{\lambda\tau}({\ }^{(2)}g_{\alpha\beta})=
{\ }^{(2)}{\cal T}_{\lambda\tau}\left(
{\ }^{(1)}g_{\alpha\beta}
\right)\ ,
\end{equation}
the
left hand side of the second order equations is identical in form to
that of the first order equation. The only difference is that the
perturbations with index $n=1$ of first order theory are replaced by
those with index $n=2$. Since the $L_{\lambda\tau}$
operator is  spherically symmetric for a Schwarzschild background,
and linear, $L_{\lambda\tau}({\ }^{(2)}g_{\alpha\beta})$
can be decomposed into multipoles, and the $\ell\,m$ multipole of 
$L_{\lambda\tau}$
will involve only that part of ${\ }^{(2)}g_{\alpha\beta}$ with the same
$\ell m$, and the same parity.

The right hand side of 
(\ref{secordeq}) can be expanded, as can any functions of ($\theta,\phi$),
in even and odd multipoles so, for example, we can write:
\begin{eqnarray}
{\ }^{(2)}{\cal T}_{0 0} & = & \sum_{\ell,m}
{\ }^{(2)}{\cal A}^{(\ell,m)} Y^\ell_m \\
{\ }^{(2)}{\cal T}_{0 2} & = & \sum_{\ell,m}
{\ }^{(2)}{\cal B}^{(\ell,m)} {\partial Y^\ell_m \over \partial
\theta}+
{\ }^{(2)}{\cal B}^{(\ell,m)\rm odd}\frac{1}{\sin\theta}\frac{\partial
Y^\ell_m}{\partial\phi}\nonumber\ ,
\end{eqnarray}
and so forth. The multipoles of 
${\ }^{(2)}{\cal T}_{0 0}$, for example, are found
from 
\begin{equation}
{\ }^{(2)}{\cal A}^{(\ell,m)}=\int {\ }^{(2)}{\cal T}_{0 0}
\left( Y^\ell_m\right)^*
d\cos\theta d\phi \ .
\end{equation}
It is important to understand that such a multipole decomposition can
always be made, but particular $\ell m$ multipole of ${\ }^{(2)}{\cal
T}_{\lambda\tau}$ will not be directly related to the same $\ell m$
multipole of the first order perturbations ${\
}^{(1)}g_{\alpha\beta}$. The nonlinear dependence of ${\ }^{(2)}{\cal
T}_{\lambda\tau}$ on the first order perturbations will mix multipoles
and parities. Thus, for example, the $\ell=2$ second order
perturbations will be driven by terms on the right hand side coming
from the product of first order perturbations with $\ell=2$ and those
with $\ell=4$. Even parity second order perturbations will be driven
by the products of odd parity first order perturbations, etc.

As an example of this mixing, Cunningham {\em et al.} \cite{cpm}
considered the collapse of a rotating relativistic stellar model, to
second order in the rate of rotation.  The star's first order
perturbation is the conserved $\ell=1$, odd parity, angular momentum
perturbation.  To second order, this rotation drives an even parity
$\ell=2$ radiatable perturbation.

The detailed expressions for the projection of the multipole
components of expressions that are quadratic in multipole expansions
(as are the ${\ }^{(2)}{\cal T}_{\lambda\tau}$) is tedious, though
straightforward \cite{zerilli2}, and will be omitted here. The manner
in which it is carried out depends on the details of a perturbation
problem. In some instances the multipole projections might best be
carried out numerically.  In what follows we shall suppose that the
necessary multipole projections have been carried out.

We shall make additional assumptions; we shall consider only the even
parity $\ell=2$ second order multipole. One justification for this is
that it will simplify rather lengthy expressions and greatly simplify
their description. This will help the presentation to focus on
important basic issues about higher order perturbation theory with
minimal distraction from a minor complication. The generalization from
$\ell=2$ to any other multipole is absolutely straightforward.  We
also choose to focus on the quadrupole for a practical
reason. Gravitational wave generation seem almost always to be
dominated by quadrupole radiation, even for sources in which the usual
arguments (slow-motion) for quadrupole dominance do not apply.  The
justification for the even parity analysis, is that it is a more
difficult system to work with. By describing the even parity second
order problem we believe we are laying the foundation for a reader to
make a similar (but simpler) extension from first order to second
order odd parity analysis.

Our next step for second order analysis is to perform a {\em purely
second order} gauge transformation that sets $ {\ }^{(2)}h_0={\
}^{(2)}h_1={\ }^{(2)}G=0 $ .  The second order gauge functions $ {\
}^{(2)}\xi^\mu $ , or equivalently $ {\ }^{(2)}A_0,{\ }^{(2)}A_1,{\
}^{(2)}A_2, $ needed are analogous to their first order counterparts,
since all changes are taking place purely at the second
order. Equations (\ref{firstxi}) -- (\ref{eventrans4}) hold true in
the second order as well as first; only the indices ``(1)'' must be
changed to ``(2)''.

At this point the range of gauge possibilities is wide, and potentially
confusing. (i) We could start in some arbitrary gauge and transform
{\em only} the second order perturbations to the RW gauge. That is we could
choose to set 
$
{\ }^{(2)}h_0={\ }^{(2)}h_1={\ }^{(2)}G=0
$,
but not necessarily 
$
{\ }^{(1)}h_0={\ }^{(1)}h_1={\ }^{(1)}G=0
$.
We would then be in a second order, but not first order, RW gauge.
(ii) We could start from an arbitrary gauge and use a first order
transformation (carried out to at least second order, of
course) to impose the conditions
$
{\ }^{(1)}h_0={\ }^{(1)}h_1={\ }^{(1)}G=0
$
but not necessarily the conditions
$
{\ }^{(2)}h_0={\ }^{(2)}h_1={\ }^{(2)}G=0
$. We would then be in a first order RW gauge, but not a second
order RW gauge. (iii) We could choose to make a first order transformation
(carried out at least to second order) to a first order RW gauge, 
followed by a purely second order transformation (which does not affect
the first order gauge) to a second order RW gauge. We would then 
have
$
h_0=h_1=G=0
$
to both first and second order, and would be in a first and second order
RW gauge.

It is easy to overlook some of the subtleties hidden in the nonlinear
interactions of the first and second order gauge transformations. One
should notice, for example, that the second order gauge functions
$
{\ }^{(2)}\xi^\mu
$
used to set
$
{\ }^{(2)}h_0={\ }^{(2)}h_1={\ }^{(2)}G=0
$
will  depend on whether we have {\em first} transformed
to the first order RW gauge (because such a transformation changes
the metric to second order).
Since the non-vanishing second order RW parts
$
{\ }^{(2)}H_0^{\rm RW}, {\ }^{(2)}H_1^{\rm RW}, 
{\ }^{(2)}H_2^{\rm RW}, {\ }^{(2)}K^{\rm RW}, 
$
depend on $
{\ }^{(2)}\xi^\mu
$,
we should keep in mind that there are not {\em unique} second
order RW perturbations. The second order RW perturbations, depend
on the first order gauge.

In a second order RW gauge, the second order Einstein equations 
(\ref{secordeq})
consist of seven equations linear in the second order functions ${\
}^{(2)}H_1^{\rm RW}$, ${\ }^{(2)}H_0^{\rm RW}$, ${\ }^{(2)}H_2^{\rm
RW}$, and ${\ }^{(2)}K^{\rm RW}$, but quadratic in the first order
functions ${\ }^{(1)}H_1^{\rm RW}$, ${\ }{(1)}H_0^{\rm RW}$, and ${\
}^{(1)}K^{\rm RW}$. One of these equations is
\begin{equation}\label{sech0h2}
{\ }^{(2)}H_2^{\rm RW}={\ }^{(2)}H_0^{\rm RW}+{\cal S}_{\rm diff}
\end{equation}
where ${\cal S}_{\rm diff}$ is quadratic in the first order
perturbations. Since the first order problem is solved independently,
${\cal S}_{\rm diff}$ can be thought of as a known ``source,'' and we
shall use the term ``source'' to refer below to similar expressions
quadratic in first order perturbations.

It is important to understand that the relation (\ref{sech0h2}) exists
in a second order RW gauge, whether or not we are using a {\em first}
order RW gauge, but the source term ${\cal S}_{\rm diff}$ will be
different (it will have a different numerical value at a given
coordinate location $t,r$) depending on the gauge choice that has been
made at first order. When it is important to emphasize the first order gauge
choice that was made in computing source terms like 
${\cal S}_{\rm diff}$, we will use a superscript to indicate the first order
gauge, so that for example 
${\cal S}_{\rm diff}^{\rm RW}$
indicates that a first order RW gauge was used, and
${\cal S}_{\rm diff}^{\rm AF}$
indicates a first order asymptotically flat gauge.

In whatever first order gauge, (\ref{sech0h2}) is the second order
 equivalent of the first order relationship ${\ }^{(1)}H_2^{\rm RW}={\
 }^{(1)}H_0^{\rm RW}$, and can be used to eliminate ${\
 }^{(2)}H_2^{\rm RW}$ in the remaining equations, as was done in the
 first order case. Note that unlike the first order case, we do not
 now introduce a symbol ${\ }^{(2)}H^{\rm RW}$ to represent both ${\
 }^{(2)}H_2^{\rm RW}$ and ${\ }^{(2)}H_0^{\rm RW}$, since these second
 order quantities are not equal.
As in the first order case, the Einstein equations consist of 
 two sets of three equations for ${\ }^{(2)}H_1^{\rm RW}$, ${\
 }^{(2)}H_0^{\rm RW}$, and ${\ }^{(2)}K^{\rm RW}$, one set containing
 only derivatives of first order in $r$, and the other with
 derivatives of second order in $r$. The functions ${\ }^{(2)}H_1^{\rm
 RW}$, ${\ }^{(2)}H_0^{\rm RW}$, and ${\ }^{(2)}K^{\rm RW}$ appear
 linearly in these equations, which have the same form, with the same
 coefficients, as for the system for the corresponding first order
 functions ${\ }^{(1)}H_1^{\rm RW}$, ${\ }{(1)}H_0^{\rm RW}$, and ${\
 }^{(1)}K^{\rm RW}$, but now with ``source'' terms quadratic in 
first order perturbations (and dependent on the first order gauge choice).
 The equations of first order in $r$ derivatives are 
\begin{eqnarray}
\frac{\partial^2 {\ }^{(2)}K^{\rm RW}} {\partial r \partial t} &=&
\frac{1}{r}{\partial {\ }^{(2)}H_0^{\rm RW} \over \partial t}+\frac{3}{r^2}
{\ }^{(2)}H_1^{\rm RW}-\frac{r-3M}{r (r-2M)}{\partial {\ }^{(2)}K^{\rm
RW} \over \partial t} +{\cal S}_K
\label{K2rt}\\
\frac{\partial^2 {\ }^{(2)}H_0^{\rm RW}}{\partial r \partial t} & = &
\frac{ r}{r-2M}{\partial^2 {\ }^{(2)}H_1^{\rm RW} \over \partial
t^2}+\frac{r-4M}{r(r-2M)}{\partial {\ }^{(2)}H_0^{\rm RW} \over
\partial t}  \nonumber \\
& &+ \frac{3M-r}{r(r-2M)}{\partial {\ }^{(2)}K^{\rm RW} \over \partial
t}+\frac{3}{r^2}
{\ }^{(2)}H_1^{\rm RW}  +{\cal S}_{H2}\\
\frac{\partial {\ }^{(2)}H_1^{\rm RW}} {\partial r} &=&
\frac{r}{r-2M}{\partial {\ }^{(2)}H_0^{\rm RW} \over \partial
t}+\frac{r}{r-2M}{\partial
{\ }^{(2)}K^{\rm RW} \over \partial t} \nonumber\\
&&\ \ \ \ \ -2\frac{M}{r(r-2M)} {\ }^{(2)}H_1^{\rm RW} + {\cal S}_{H1}
\label{H12r}\ ,
\end{eqnarray}
%%%%%%%%%%%%%%%%%%%%%
where ${\cal S}_K$ and ${\cal S}_{H2}$ are ``source terms,'' 
quadratic in 
first order perturbations, and in the $r$ and $t$ derivatives
of first order perturbations.

As in the first order case, the second order RW perturbations must
satisfy the remaining Einstein equations, three equations with second
order derivatives in $r$. These equations are analogous to the corresponding
first order equations, but the second order equations contain source terms
quadratic in the first order perturbations. The procedure that was used
to simplify the first order system  works also in the second
order case. 
For each arrangement of substitutions a different source term appears, but
one can show that the source terms are equal if the first order perturbations
that appear in the source terms satisfy the first order perturbation
equations.
 This means that the equations that must be solved are 
three equations of first order in $r$ derivatives, plus an ``algebraic
identity,'' of the form
\begin{eqnarray}
\label{algid2}
 \frac{r^2}{r-2M}\frac{\partial^3 {\ }^{(2)}K^{\rm RW}}{\partial t^3}-
 \frac{\partial^2
 {\ }^{(2)}H_1^{\rm RW}}{\partial t^2}&+&\frac{2r^2-3
Mr-3M^2}{r^2(r-2M)}\frac{\partial {\ }^{(2)}K^{\rm RW}}{\partial
t}-\frac{3M}{r^3} {\ }^{(2)}H_1^{\rm RW} \nonumber \\
-\frac{2r+3M}{r^2}\frac{\partial {\ }^{(2)}H_0^{\rm RW}}{\partial t}&+&
{\cal S}_{AI}  =  0\ ,
\end{eqnarray}
where ${\cal S}_{AI}$ is a ``source'' term, quadratic in 
the first order perturbations.
%${\ }^{(2)}H_1^{\rm RW}$
In the same manner as in the first order case, the next step is to use
the ``algebraic identity'' (\ref{algid2}) to eliminate
${\ }^{(2)}H_0^{\rm RW},_t$ in (\ref{K2rt}) and (\ref{H12r}), and
thereby to reduce the system  to a set of two coupled linear partial
differential equations (with ``sources'') for ${\ }^{(2)}K^{\rm RW}$
and ${\ }^{(2)}H_1^{\rm RW}$.
It is 
immediate to check that the part linear 
in ${\ }^{(2)}K^{\rm RW},_t$ and ${\ }^{(2)}H_1^{\rm RW}$
has exactly the same form
as the corresponding system for ${\ }^{(1)}K^{\rm RW}$ and ${\
}^{(1)}H_1^{\rm RW}$ in the first order perturbation case, as indeed
it must. We may then, as in the first order perturbation treatment,
introduce a ``diagonalization'' procedure by defining the functions
${\ }^{(2)}\chi$ and ${\ }^{(2)}\widehat{R}$, such that
\begin{equation}
\frac{\partial {\ }^{(2)}K^{\rm RW} }{ \partial t}=f(r) 
 {\ }^{(2)}\chi
+g(r) \frac{\partial {\ }^{(2)}\widehat{R} }{ \partial t} \;\;,\;\;
{\ }^{(2)}H_1^{\rm RW}
= h(r)
{\ }^{(2)}\chi
+k(r)
{\ }^{(2)}\widehat{R} 
\label{Zer02}
\end{equation}
where $f$, $g$, $h$ and $k$ are the same as in (\ref{handk}). Using this
transformation we may check that the system 
for  ${\ }^{(2)}K^{\rm RW},_t$ and 
${\ }^{(2)}H_1^{\rm RW}$
is equivalent to
\begin{equation}
\frac{\partial {\ }^{(2)}\chi} { \partial r^*} =
 {\ }^{(2)}\widehat{R}
+ \widehat{\cal S}_{ZK} \;\;,\;\;
\frac{\partial {\ }^{(2)}\widehat{R}} {\partial r^*} =
\left[V(r^*) + \frac{\partial^2 } {\partial t^2} \right]{\ }^{(2)}\chi
+\widehat{\cal S}_{ZR}
\label{Zer12}\ .
\end{equation}
This, again, implies that ${\ }^{(2)}\chi$ satisfies an equation of the  form,
\begin{equation}
{\partial^2  \over \partial {r^*}^2}  {\ }^{(2)}\chi - 
{\partial^2  \over \partial {t}^2}  {\ }^{(2)}\chi 
 -V(r^*) 
{\ }^{(2)}\chi  + {\cal S}_Z =0
\label{Zer22}
\end{equation}
where ${\cal S}_Z$ is the  ``source'' term for this second 
order Zerilli equation.

The procedure for specifying the Cauchy data for (\ref{Zer22}) is
patterned on that for first order calculations.  From our family of
initial value solutions, we suppose that we have the 3-metric
$\gamma_{ij}$ and the extrinsic curvature $K_{ij}$ to second order as
well as first. We assume, for convenience, that the initial value
solutions is in a spacetime gauge with ${\ }^{(2)}\widetilde{h}_0 ={\
}^{(2)}\widetilde{H}_1 ={\ }^{(2)}\widetilde{H}_0=0$. With
(\ref{Kgamdot}) we then have the initial values of the perturbation
functions ${\ }^{(2)}\widetilde{K}$, ${\ }^{(2)}\widetilde{G}$, ${\
}^{(2)}\widetilde{H}_2$, ${\ }^{(2)}\widetilde{h}_1$, and their first
time derivatives.  We next use the second order equivalent of
(\ref{eventrans1}) --- (\ref{eventrans4}). It should be recalled that
these equations are the result of a purely second order gauge transformation;
the second order version of (\ref{eventrans1}) --- (\ref{eventrans4})
is precisely the same as the first order, with ``1'' superscripts
replaced by ``2''. There are no additional terms quadratic in first
order perturbations.  These relations allow us to compute the initial
values of ${\ }^{(2)}K^{(\rm RW)}$,${\ }^{(2)}\dot{K}^{(\rm RW)}$ and
${\ }^{(2)}H_1^{(\rm RW)}$.
The next step, following the first order pattern, is 
to use the second order Einstein equations
\begin{equation}\label{Hdoteq2}
r\frac{\partial {\ }^{(2)}H_1^{RW}  }{\partial t}
=\frac{2M}{r}{\ }^{(2)}H_2^{RW}
+\left(r-2M\right)
\left(
\frac{\partial {\ }^{(2)}H_2^{RW}  }{\partial r}
-\frac{\partial {\ }^{(2)}K^{RW}  }{\partial r}
\right)+{\cal S}_{H1t}^{\rm RW}
\end{equation}
\begin{displaymath}
\frac{r^3}{r-2M}\frac{\partial^2 {\ }^{(2)}K^{RW}  }{\partial t^2}
=
r(r-2M)\frac{\partial^2 {\ }^{(2)}K^{RW}  }{\partial r^2}
+
2(2r-3M)\frac{\partial {\ }^{(2)}K^{RW}  }{\partial r}
\end{displaymath}\begin{equation}
\label{Kdotdoteq2}
-
2(r-2M)\frac{\partial {\ }^{(2)}H_2^{RW}  }{\partial r}
+
2r\frac{\partial {\ }^{(2)}H_1^{RW}  }{\partial t}
+
(\ell-1)(\ell+2){\ }^{(2)}K^{RW}
-2{\ }^{(2)}H_2^{RW}+{\cal S}_{Ktt}^{\rm RW},
\end{equation}
the second order equivalents of 
(\ref{Hdoteq}) and
(\ref{Kdotdoteq}), to find the initial values of  
${\ }^{(2)}\dot{H}_1^{RW}$
and
${\ }^{(2)}\ddot{K}^{RW}$.
With these and the definitions in 
(\ref{Zer02})
the determination of ${\ }^{(2)}\chi$ and ${\ }^{(2)}\dot{\chi}$\,, at $t=0$,
is complete.

\subsection{Radiation}

Above we have given a description of how to solve for the second order
Zerilli function ${\ }^{(2)}\chi$. It is important to understand that
this second order perturbation cannot be considered as a second order
correction to ${\ }^{(1)}\chi$.  Rather, we must transform to
coordinates which are asymptotically flat (AF) at least to second
order in the perturbations. In this AF coordinates system, the
dominant terms as $r\rightarrow\infty$ are the transverse $K$ and $G$
terms, which fall off as $1/r$. Most important, in the AF coordinate
system the $K$ and $G$ terms give us the intensity of the outgoing
radiation.  The amplitude of the outgoing radiation, correct to first
and second order in perturbation theory, therefore, is given by $K$
and $G$ correct to first and second order, in a gauge that is AF to
first and second order.

In the problem of finding the radiation to second order there are
calculational and conceptual problems that arise that are not present
in the purely first order problem. In discussing these we start by
pointing out how, in principle, the second order radiation could be
computed in a straightforward way. We could, {\em ab initio}, invoke an AF
gauge. (The specification of a unique AF gauge is not itself well
defined, but that is not the issue here.)  In this AF gauge we could
write out the Einstein equations to first and to second order. 
The
source terms ${\ }^{(2)}{\cal T}_{\alpha\beta}$ would have a form
different from the form we use, based on a ${\ }^{(2)}{\cal
T}_{\alpha\beta}$ computed in a RW gauge. And the various source 
terms 
$
{\cal S}_K^{\rm AF}$,
${\cal S}_{H1}^{\rm AF}$,
${\cal S}_{H2}^{\rm AF}$,
${\cal S}_{AI}^{\rm AF},{\cal S}_{Z}^{\rm AF}$
would have different values than the source terms
$
{\cal S}_K^{\rm RW}$,
${\cal S}_{H1}^{\rm RW}$,
${\cal S}_{H2}^{\rm RW}$,
${\cal S}_{AI}^{\rm RW},{\cal S}_{Z}^{\rm RW}$
. The resulting 
${\ }^{(2)}\chi$
is a combination of second order RW perturbation functions, and
as in the first order case, the relationship to $K$ and $G$
in a second order AF gauge can be found 
by a  second order equivalent of a first order procedure.
In practice one cannot use a first order AF gauge. 
The enormous complexity of the source term underscores the
need for a simple, as well as definitive choice of gauge.
Our approach is to 
work in a first order RW gauge and to compute 
$
{\cal S}_K^{\rm RW}$,
${\cal S}_{H1}^{\rm RW}$,
${\cal S}_{H2}^{\rm RW}$,
${\cal S}_{AI}^{\rm RW},{\cal S}_{Z}^{\rm RW}$.

We will then, in principle compute ${\ }^{(2)}\chi$ in a first order
RW gauge. In practice, there is a new difficulty not present in 
the first order problem.
When the first order RW gauge is used, the source term ${\cal
S}_{Z}^{\rm RW}$ in the second order Zerilli equation diverges at
large $r$, and any solution ${\ }^{(2)}\chi$ must diverge. This is of
course a gauge effect (the effect of computing the sources in a first
order RW gauge) and not an indication of a physical divergence. But as
a practical matter we cannot do computations with a divergent 
quantity. 
In practice, therefore, we need to modify the problem to make it
numerically tractable. This is possible since one does not spoil the
invariance of ${\
}^{(2)}\chi$ under pure second order gauge transformations by adding
pieces
that are quadratic in the first order perturbations. This highlights
the fact that ${\
}^{(2)}\chi$ is far from a unique quantity, in fact there is an
infinite family of possible candidates for a ``second order Zerilli
function''. We therefore add a known term to ${\
}^{(2)}\chi$ to cancel the divergent large-$r$ behavior and to define
a ``renormalized''  
${\ }^{(2)}\chi_n$ by
\begin{equation}\label{Gammadef}
{\ }^{(2)}\chi_n={\ }^{(2)}\chi+\Gamma
\end{equation}
which satisfies
\begin{eqnarray}\label{renormd}
\frac{\partial^2
{\ }^{(2)}\chi_n
}{\partial r^{*2}}-\frac{\partial^2
{\ }^{(2)}\chi_n
}{\partial t^{2}}-V(r^*){\ }^{(2)}\chi_n=-{\ }^{(2)}{\cal S}_{Zn}\nonumber\\
=-\frac{\partial^2
\Gamma
}{\partial r^{*2}}+\frac{\partial^2
\Gamma
}{\partial t^{2}}+V(r^*)\Gamma-{\ }^{(2)}{\cal S}_{Z}
\end{eqnarray}
Here $\Gamma$ is a {\em known} expression quadratic in the (RW gauge) first
order perturbations constructed so that the extra terms on the
right hand side of (\ref{renormd}) cancel the dominant large-$r$
behavior. Since $\Gamma$ is known, numerically solving for ${\
}^{(2)}\chi_n$ is equivalent to solving for ${\ }^{(2)}\chi$.

The explicit forms of source terms depend on the details of the problem
and, in particular, on how the various first order multipoles contribute.
We do not, therefore, give explicit general expressions for the
source terms. 
${\cal S}_K^{\rm RW}$,
${\cal S}_{H1}^{\rm RW}$,
${\cal S}_{H2}^{\rm RW}$,
${\cal S}_{AI}^{\rm RW}$,
${\cal S}_{Z}^{\rm RW}$,
${\cal S}_{Zn}^{\rm RW}$.
Rather, as an example, we consider the case that contributions to the 
second order even parity quadrupole come exclusively from the 
first order even parity quadrupole. This specialization, in fact, 
applies to the particular configurations to which second order analysis
has already been applied \cite{gnppprl}-- \cite{gnppboost}.
In this case, the renormalization is accomplished with 
\begin{equation} 
\chi^{(2)}_n = \chi^{(2)}
-{1 \over 7} \sqrt{{\pi \over 5}} \left[
{ r^2 \over 2r+3M } {\ }^{(1)}K^{\rm RW} {\ }^{(1)}K^{\rm RW},_t
+ ({\ }^{(1)}K^{\rm RW})^2 \right]
\end{equation}
and the resulting source ${\cal S}_{Zn}^{\rm RW}
$, a quadratic expression in the first order perturbation, and some of
their space and time derivatives, may be written entirely in terms of
${\ }^{(1)}\psi$, using (\ref{Kdepsi}) -- (\ref{Hdepsi}). The explicit
expression 
\footnote{As correctly pointed out by Davies  \cite{Davies}, there is a
misprint in reference  \cite{gnppcqg}, the term involving
$\dot{\psi}^2$ should be multiplied by $1/\mu^3$ (if not it would not
be even dimensionally right). For the historical record, the misprint
occurred in typing the manuscript, i.e. the correct formula was used
in the numerical results derived with this formalism,
e.g. in reference  \cite{gnppprl}.} is given in Eq. (18), in
 \cite{gnppcqg}.

The process of extracting information about radiation from 
the solution for
$
\chi^{(2)}_n(t,r)
$ 
follows the general pattern of the first order problem. 
One starts by writing the solution 
for
$
{\ }^{(2)}\chi_n(t,r)
$
in the form 
\begin{equation}\label{F2exp}
{\ }^{(2)}\chi_n(t,r) = {\ }^{(2)}F_a(t-r^*) + {\ }^{(2)}F_b(t-r^*)/r 
+{\ }^{(2)}F_c(t-r^*)/r^2 + {\cal O}(1/r^3)\ .
\end{equation}
From the second order Zerilli equation (\ref{Zer22}), one can show
that ${\ }^{(2)}F_b$, ${\ }^{(2)}F_c$, etc. are determined once ${\
}^{(2)}F_a$ and the first order RW perturbations are known.  For the
special (but important) case that the only relevant first order
perturbations are the even parity axisymmetric quadrupole
perturbations,  we have

\begin{eqnarray}
{\ }^{(2)}F_b(t ) & = & 3 \int^t {\ }^{(2)}F_a  dt' + {2 \over 7}\sqrt{
{\pi \over 5}}\left\{ {1 \over 3} {d{\ }^{(1)}F  \over d t}
{d^4{\ }^{(1)}F  \over d t^4} + {d^2{\ }^{(1)}F  \over d
t^2}{d^3{\ }^{(1)}F  \over d t^3} \right. \nonumber \\
&& \left. +{1 \over 3} \int^t \left({d^3{\ }^{(1)}F  \over d t'^3}
\right)^2 dt' +{M \over 6} \left[ {d^2{\ }^{(1)}F  \over d
t^2}{d^4{\ }^{(1)}F  \over d t^4}
 - \left({d^3{\ }^{(1)}F  \over d t^3}\right)\right] \right\} .
\end{eqnarray}
where 
$
{\ }^{(1)}F
$ is the asymptotic function introduced in (\ref{Fexpan}), and we may
obtain similar expressions for ${\ }^{(2)}F_c(t )$, and higher order
coefficients.

With this form for 
$
\chi^{(2)}(t,r)
$, one then uses the second order system in (\ref{K2rt})---
(\ref{Zer02}), to solve for the second order RW functions.  Those
functions will, of course, diverge at large $r$ due to  now familiar
gauge effects.
This solution gives results,
in terms of
the asymptotic functions ${\ }^{(2)}F_a$, ${\ }^{(2)}F_b$, and ${\
}^{(2)}F_c$, for the functions
${\ }^{(2)}K^{\rm RW}$,
${\ }^{(2)}H_0^{\rm RW}$,
${\ }^{(2)}H_1^{\rm RW}$,
${\ }^{(2)}H_2^{\rm RW}$,
which are perturbations in a gauge that is RW to both first and 
second order. To extract information about outgoing radiation
we must now make two asymptotic gauge transformations. First we must 
transform from the first order RW gauge to an ``intermediate'' 
gauge, the result of a first order transformation that makes our
coordinates system AF to first, but not to second, order. We have
already discussed this transformation in Sec.\,IIIB, in particular in
equations (\ref{alpha1}) --- (\ref{firstgavec}), and found the
asymptotic form of the gauge transformation functions
${\ }^{(1)}\alpha_0^{\ell,m}$,
${\ }^{(1)}\alpha_1^{\ell,m}$,
${\ }^{(1)}\alpha_2^{\ell,m}$,
for each multipole. In Sec.\,IIIB we were 
concerned only with the effect of this transformation on first order 
perturbations. Here we are concerned with the effect 
of the transformations on second
order perturbations, and to second order, the effect of the transformation 
will be quadratic
in the gauge functions
${\ }^{(1)}\alpha_0^{\ell,m}$,
${\ }^{(1)}\alpha_1^{\ell,m}$,
${\ }^{(1)}\alpha_2^{\ell,m}$,
and will have the form, for example
\begin{equation}\label{alphexample}
{\ }^{(2)}H_0^{\rm INT}={\ }^{(2)}H_0^{\rm RW}+
{\rm quad}\ ,
\end{equation}
where ``INT'' represents the intermediate gauge, and ``quad'' 
stands for some expression quadratic in the first order terms,
either in products of the ${\ }^{(1)}\alpha$ gauge functions, or 
products of ${\ }^{(1)}\alpha$ gauge functions with first order
perturbations.
For the first order effects all equations were linear, and each multipole
of the ${\ }^{(1)}\alpha^{(\ell,m)}$ gauge 
functions had an effect only on the same multipole
of the metric perturbations. In the second order effects, the multipoles
mix, and no useful general expression can be given for ``quad.'' 
In our most familiar case, when the only first order perturbations of
concern are the even parity, axisymmetric, quadrupole perturbations, the
explicit gauge transformation is, for example,
\begin{displaymath}
{\ }^{(2)}K^{\rm INT}={\ }^{(2)}K^{\rm RW}-\frac{4}{7r} 
\sqrt{{\pi \over 5}} \left\{
\left[
-r{\ }^{(1)}\alpha_0{\ }^{(1)}K^{\rm RW},_t-{\ }^{(1)}\alpha_1
{\ }^{(1)}K^{\rm RW},_r 
\right. \right.
\end{displaymath}\begin{displaymath}\left.
+9r{\ }^{(1)}\alpha_2 {\ }^{(1)}K^{\rm RW} 
-2{\ }^{(1)}\alpha_1 {\ }^{(1)}K^{\rm RW}\right]
+\frac{1}{r^2(r-2M)}\left[-2Mr{\ }^{(1)}\alpha_1^2
-6r^2{\ }^{(1)}\alpha_0^2
\right.
\end{displaymath}\begin{displaymath}
\left.
+7r^2{\ }^{(1)}\alpha_1^2
+24r^4{\ }^{(1)}\alpha_2^2
-24M^2{\ }^{(1)}\alpha_0^2
+36Mr^2{\ }^{(1)}\alpha_1{\ }^{(1)}\alpha_2
-18r^3{\ }^{(1)}\alpha_1{\ }^{(1)}\alpha_2
\right.
\end{displaymath}\begin{displaymath}
\left. 
-48Mr^3{\ }^{(1)}\alpha_2^2
+24Mr{\ }^{(1)}\alpha_0^2
-4Mr^2{\ }^{(1)}\alpha_1 {\ }^{(1)}\alpha_1,_r
-4Mr^2{\ }^{(1)}\alpha_0{\ }^{(1)}\alpha_1,_t
\right.
\end{displaymath}
\begin{equation}\label{alphexplicit}
\left. \left.
+2r^3{\ }^{(1)}\alpha_1{\ }^{(1)}\alpha_1,_r
+2r^3{\ }^{(1)}\alpha_0{\ }^{(1)}\alpha_1,_t
\right]\  \right\} .
\end{equation}
With such expressions, and with the already established asymptotic forms
of the gauge functions given in
(\ref{avsF}), or 
(\ref{firstgavec}), we now have the relationship
between the asymptotic form of ${\ }^{(2)}\chi$ and of the metric
perturbations in the intermediate gauge. Equivalently, we have the
asymptotic forms of
${\ }^{(2)}H_0^{\rm INT}$,
${\ }^{(2)}H_1^{\rm INT}$,
${\ }^{(2)}H_2^{\rm INT}$,
${\ }^{(2)}h_0^{\rm INT}$,
${\ }^{(2)}h_1^{\rm INT}$,
${\ }^{(2)}K^{\rm INT}$,
${\ }^{(2)}G^{\rm INT}$,
in terms of 
${\ }^{(2)}F_a(t-r*)$, and functions related to it.

The second step of our process is to perform a purely second order
transformation  to take us from the intermediate gauge to a gauge
that is AF to both first and second order. We denote the gauge functions
that implement this (even parity, quadrupolar, axisymmetric) transformation 
with
${\ }^{(2)}\alpha$, that is,
our second order gauge transformation uses
$$
\xi^{(2)}{}^0 = {\ }^{(2)}\alpha_0(t,r) Y_\ell^m \;\;\;,\;\;\; \xi^{(2)}{}^1 =
{\ }^{(2)}\alpha_1(t,r) Y_\ell^m
$$
\begin{equation}
\xi^{(2)}{}^2 = {\ }^{(2)}\alpha_2(t,r) {\partial \over \partial \theta} Y_\ell^m
\;\;\;,\;\;\;
\xi^{(2)}{}^3 = {\ }^{(2)}\alpha_2(t,r)  {1 \over \sin^2 \theta} {\partial \over
\partial \phi}Y_\ell^m
\end{equation}
for $\ell,m=2,0$.
These gauge functions, from RW to AF,  have the asymptotic form
\begin{eqnarray}
{\ }^{(2)}\alpha_0
& = & r {\ }^{(2)}\alpha_{0a}(t-r^*) + {\ }^{(2)}\alpha_{0b}(t-r^*) +
{\ }^{(2)}\alpha_{0c}(t-r^*)/r + \cdots \nonumber\\
{\ }^{(2)}\alpha_1 & = & r {\ }^{(2)}\alpha_{1a}(t-r^*) +
{\ }^{(2)}\alpha_{1b}(t-r^*) + {\ }^{(2)}\alpha_{1c}(t-r^*)/r +\cdots
\\
{\ }^{(2)}\alpha_2   & = & {\ }^{(2)}\alpha_{2a}(t-r^*)/r +
{\ }^{(2)}\alpha_{2b}(t-r^*)/r^2 + {\ }^{(2)}\alpha_{2c}(t-r^*)/r^3 +
\cdots\ .
\nonumber
\end{eqnarray}
In the special case that the only first order perturbations that
contribute to the second order quadrupole equations are the even 
parity axisymmetric quadrupole perturbations, the coefficient
functions are given by
\begin{eqnarray} 
{\ }^{(2)}\alpha_{0a}(t) & = & -{1\over 2} {\ }^{(2)}F_a(t)  + {1 \over
63} \sqrt{{\pi \over 5}}
 \left( {\ }^{(1)}F'''(t) \right)^2  \\
{\ }^{(2)}\alpha_{1a}(t) & = & -{1\over 2} {\ }^{(2)}F_a(t)  + {1 \over
63} \sqrt{{\pi \over 5}}
 \left( {\ }^{(1)}F'''(t) \right)^2  \\
{\ }^{(2)}\alpha_{2a}(t) & = & -{1\over 2} \int^t{\ }^{(2)}F_a(t') dt'
+ {1 \over 126}
\sqrt{{\pi \over 5}} 
 {\ }^{(1)}F''(t)  {\ }^{(1)}F'''(t)   \\
{\ }^{(2)}\alpha_{0b}(t) & = & -\int^t{\ }^{(2)}F_a(t') dt'
-M {\ }^{(2)}F_a(t) -{M \over 63} \sqrt{{\pi \over 5}} 
 \left( {\ }^{(1)}F'''(t) \right)^2  \\
{\ }^{(2)}\alpha_{1b}(t) & = & -{3\over 2} \int^t{\ }^{(2)}F_a(t') dt'
  \\
{\ }^{(2)}\alpha_{2b}(t) & = & - \int^t \left[\int^{t'} {\ }^{(2)}F_a(t'') dt''
\right] dt'  - {2 \over 63} \sqrt{{\pi \over 5}}
  {\ }^{(1)}F'(t)  {\ }^{(1)}F'''(t) 
+ {1 \over 126} \sqrt{{\pi \over 5}}
 \left( {\ }^{(1)}F''(t) \right)^2   
\end{eqnarray} 
The function ${\ }^{(1)}F$
is known from the solution to the first order problem, and 
${\ }^{(2)}F_a$ is defined in 
({\ref{F2exp})
as the asymptotic part of the second order Zerilli function.

The result of the two step gauge transformation is expressions, in terms 
of ${\ }^{(2)}F_a (t-r^*)$, and of first order functions, for the asymptotic
second order metric perturbations in a (first and second order) AF gauge.
Among these relations, we have
${\ }^{(2)}G^{\rm AF}$
and
${\ }^{(2)}K^{\rm AF}$ that carry information about the radiation. For the
special case of  only  even parity axisymmetric
quadrupole first order perturbations the results is
\begin{eqnarray}
{\partial {\ }^{(2)}{G}^{\rm AF} \over \partial t}
 & = & {1 \over r} \left\{ {\ }^{(2)}F_a (t-r^*) 
+{2 \over 63} \sqrt{{\pi \over 5}}
{\partial \over \partial t} \left[  {\ }^{(1)}F''(t-r^*)
{\ }^{(1)}F'''(t-r^*) \right] \right\} + O(r^{-2}) \\
{\partial {\ }^{(2)}{K}^{\rm AF} \over \partial t}
 & = & {3 \over r} \left\{ {\ }^{(2)}F_a (t-r^*) 
+{2 \over 63}\sqrt{{\pi \over 5}}
{\partial \over \partial t} \left[  {\ }^{(1)}F''(t-r^*)
{\ }^{(1)}F'''(t-r^*) \right] \right\} + O(r^{-2})
 \ .
\end{eqnarray}
In terms of the actual results 
${\ }^{(1)}\psi$
and
${\ }^{(2)}\chi_n$
of first and second order computations, these can be written.
\begin{eqnarray}
{\partial {\ }^{(2)}{G}^{\rm AF} \over \partial t}
& = & {1 \over r} \left[ {\ }^{(2)}\chi_n(t,r) 
+{2 \over 7} \sqrt{{\pi \over 5}}
{\partial \over \partial t} \left({\ }^{(1)}\psi(t,r) {\partial
{\ }^{(1)}\psi(t,r) \over \partial t} \right)
\right]
+ O(r^{-2}) \\
{\partial {\ }^{(2)}{K}^{\rm AF} \over \partial t} & = & {3 \over r}
\left[ {\ }^{(2)}\chi_n(t,r)
+{2 \over 7} \sqrt{{\pi \over 5}}
{\partial \over \partial t} \left({\ }^{(1)}\psi(t,r) {\partial
{\ }^{(1)}\psi(t,r) \over \partial t} \right)    \right] +
O(r^{-2})\ .
\end{eqnarray}
All information about gravitational wave energy is carried by
$G$ and $F$ 
in an AF gauge and, in (\ref{chiAF}), we have seen that 
${\ }^{(1)}\dot{G}^{\rm AF}
=
{\ }^{(1)}\dot{K}^{\rm AF}/3
={\ }^{(1)}\chi(t,r)/r\ ,
$
for $\ell=2$.
We may therefore interpret 
the expression in
brackets in ${\ }^{(2)}\dot{G}^{\rm AF}$ 
as the "second order correction to the gravitational
wave amplitude $\chi$". In particular, we have that the 
gravitational wave quadrupole power is

\begin{equation}\label{2ndPower1}
{\rm Power}=\frac{6 \pi}{25}
\left\{
{\ }^{(1)}\chi+\epsilon\left[
{\ }^{(2)}\chi
+{2 \over 7} \sqrt{{\pi \over 5}}
{\partial \over \partial t} \left({\ }^{(1)}\psi(t,r) {\partial
{\ }^{(1)}\psi(t,r) \over \partial t} \right)
\right]
\right\}^2\  .
\end{equation}
We may choose to use the expression in (\ref{2ndPower1}), or 
to keep only the terms which are explicitly second order, and
compute the energy from

\begin{equation}\label{2ndPower2}
{\rm Power}=\frac{6 \pi}{25}
\left\{
{\ }^{(1)}\chi+2 \epsilon {\ }^{(1)}\chi\left[
{\ }^{(2)}\chi
+{2 \over 7} \sqrt{{\pi \over 5}}
{\partial \over \partial t} \left({\ }^{(1)}\psi(t,r) {\partial
{\ }^{(1)}\psi(t,r) \over \partial t} \right)
\right]
\right\}\  .
\end{equation}
which differs from 
(\ref{2ndPower1}) to third order. 
For comparison with numerical work \cite{gnppprl,gnppboost}
we take the expression in curly brackets in (\ref{2ndPower1}), aside
from overall normalization, to be the gravitational wave amplitude correct to
second order. We compare radiated energy to the time integral of the 
power given in (\ref{2ndPower2}), although the equally justifiable
expression in (\ref{2ndPower1}) turned out to give better agreement 
with the numerical results used.

\subsection{Second order techniques for collisions}

The first and simplest application of second order calculations to
collisions was the analysis of the collision starting with the Misner
initial data described in (\ref{mismet}) -- (\ref{kappaeq}). The
family of spacetimes that evolves from these data is described by two
parameters, but one of them is a trivial overall scaling (say the
initial ADM mass).  The remaining parameter $\mu_0$ gives a
dimensionless measure of initial separation and this parameter (more
properly, some function of this parameter) is the basis of our
ordering of perturbations.  [See the discussion following
(\ref{kappaeq}).] The computations based on this scheme have been
presented \cite{gnppprl}, and a comparison given of numerical
relativity results, results of first order perturbation computations,
and the results of perturbation theory to second order.  The
comparison showed precisely the pattern predicted: where the results
of first order computations and of second oder computations began to
diverge (which occurred for parameter $\mu_0$ around 1.8) the results
of either order started to diverge significantly from numerical
results.  This confirmed that second order perturbation would have
told us the limiting range of perturbation results if we had not had
available the results of numerical relativity.

Subsequent applications of perturbation theory have involved
additional complications. Perturbation theory has been
applied \cite{etal} to spacetimes evolving from
Bowen-York \cite{bowenyork} initial data corresponding to two equal
mass holes which are initially moving symmetrically toward each other.
In this case we have three parameters. One is an overall scaling and
can be taken to be the initial ADM mass $M$ of the spacetime.  A
second parameter is the magnitude of the initial momentum $P$ of each
hole, and a third is some measure $L$ of the initial distance between
the holes. Aside from the overall scaling, the family of spacetimes
can be characterized by two dimensionless parameters, say $P/M$ and
$L/M$. The perturbation analysis reported in  \cite{etal} treated both
of these parameters as small.  In order to apply perturbation methods
to such a multiparameter family, it is useful to consider a curve
through the parameter space. Such a curve gives us a one parameter
family of spacetimes, and we can then apply standard methods. The
curve through the parameter space, however, is not unique. As an
example, let us suppose that we are considering initial data for
$P/M=0.3$ and $L/M=0.1$. To find perturbation results for this
example, we could consider that we are on a curve $P/M=3L/M$, and
$L/M$ is our perturbation parameter. But we could equally well treat
the spacetime as a point along the curve $P/M=30(L/M)^2$. The
perturbation analysis for close/slow (i.e., small $P$, small $L$)
perturbation theory will depend on which curve in parameter space was
chosen. The agreement with numerical relativity results will not be
equally good for the two choices.

A different sort of choice was described following (\ref{alternate}).
It was pointed out that for second order perturbation theory, one
could ``feed back'' all information about the first order
perturbations; this gives a result that differs only to third and
higher order from ``standard'' second order perturbation theory.  It
was also pointed out that procedure destroys the spherical symmetry of
the differential operators in the perturbation equations and
enormously complicates the analysis.

There is, however, a way of using updating without paying the price of
loss of symmetry, and it turns out to give an important improvement.
We can update only the information about the monopole, thereby
preserving spherical symmetry of the operators. This turns out to be
quite important in the slow/close perturbation analysis of initially
boosted holes \cite{etal,gnppboost}. The ADM mass of the spacetime
turns out to be much more sensitive to the initial momentum, than the
quadrupole deformation is. A first order change in the monopole mixes
with a first order change in the quadrupole to produce a second order
quadrupole deformation, so our second order radiation computations are
influenced by the change in the mass.  If we use the standard
approach, the rapid growth of the ADM mass with momentum is the
limiting factor that determines the disappointingly small range of
applicability of perturbation theory. To avoid this limitation we can
use a higher order estimate of the ADM mass. We have, in fact, used a
numerical (rather than perturbative) computation of the ADM mass
corresponding to a particular set of parameters $P$ and $L$. This
procedure results in an enormous improvement in the range over which
perturbation calculations can be used for radiation. The reason for
this is clear in the following: For values of $P$ large enough so that
the ADM mass increases by several hundred percent, the radiated energy
is still very small. 

This same phenomenon was found in our analysis of the radiation
generated as a single spinning Bowen-York \cite{bowenyork} hole evolves
to its final Kerr state. This problem was analyzed in the limit of
slow rotation, so that it could be treated as a perturbation of a
Schwarzschild hole \cite{gnppspin}.  The effect of the spin on the
ADM mass was again found to be much larger than the effect on
radiatable multipoles. Again, a numerical evaluation of the ADM mass
was used to extend the range of the analysis.

%%%%%%%%%%%%%%%%%%%%%%%%%%%%%%%%%%%%%%%%%%%%%%%%%%%%%%

\section{Summary}

Buried, not too deeply we hope, in the equations of the previous
section are a few general lessons that justify emphasis. We see, for
one thing, that in second order work, one relinquishes all the
simplicity of linearity that characterizes first order perturbation
calculations. Gauge transformations, in particular, have a completely
different character when one works at second order; gauge
transformations of first order change second order perturbations. In
our approach to second order perturbations, these gauge
transformations could not be avoided for a rather basic reason: One
does the mathematics in one gauge, and the physics in others.  Actual
numerical second order computations have to be carried out in a gauge
which is both convenient (i.e., minimizes the proliferation of terms)
and (unlike the ``asymptotically flat gauge'') is definitive. In our
approach that means the use of a system that satisfies the
Regge--Wheeler 
gauge conditions to both first and second order. But the
Regge--Wheeler gauge is generally not related to two other gauges that
are of importance to the problem of evolving initial data and finding
radiation. First, initial data will be found for the problem in some
``initial data'' gauge, and must be transformed to Regge--Wheeler gauge
to give us Cauchy data for evolution. Secondly, the result of second
order evolution must be transformed, asymptotically, to an
asymptotically flat gauge in order for information to be extracted
about outgoing wave amplitudes.  One must explicitly perform gauge
transformations in the process of making these calculations. And the
gauge transformations require two steps. One must, in general, perform a
first order gauge transformation carried out to second order, and next
an exclusively second order gauge transformation.

There are, in principle, other ways to proceed with second order
calculations.  It is possible, in principle, to construct expressions
which are formally gauge invariant to both first and second order. The
same formal expressions could be used for setting the Cauchy data, for
evolution, and for interpretation of the results. Such gauge invariant
expressions would, in effect, have built into them the ``gauge
invariant'' (i.e., unique) character of the Regge--Wheeler gauge. Since
the procedure for going from an arbitrary gauge to a (first and second
order) Regge--Wheeler gauge is unambiguous and local, all second order
Regge--Wheeler gauge quantities can be written as combinations of
perturbations in a general gauge.  The development of such a formalism
is underway \cite{garat}. By their gauge invariant character such
expressions would have built into them the gauge transformations, that
we perform, in a manner of speaking, as an ``external'' process. The
gauge invariant expressions will therefore be significantly more
complex in appearance, but the important issue is not so much
appearance, as suitability for computation.  Since these computations
(e.g., the evaluation of source terms) will be carried out
numerically, a gauge invariant expression may turn out be more subject
to round off error or less, and roundoff error can be a significant
problem in expressions combining many high derivative terms.

In closing, we note that second order perturbation theory has turned
 out to be a great deal more difficult than linearized theory, but
 overcoming these difficulties is motivated by the fact that second
 order calculations are a great deal easier than numerical relativity.
 What can be (speculatively) said of third and higher order
 calculations? On the one hand we suggest that the step from second
 order to third and higher order might not be as painful as the step
 from first to second order. The step up to second order required
 developing the new tools for dealing with nonlinearities.  With the
 pattern of those tools established, and with the conceptual issues
 faced, the next step up adds complexity, but, we believe, no new
 conceptual difficulties. The complexity added by each step up in
 order is considerable, but such complexity is not a crucial
 obstacle if the work is being done, as it certainly must be, by
 computers.

To balance this argument. that one should not be terrified of yet
higher order perturbation calculations, it should be asked what is to
be gained by such higher order results. The overwhelming motivation
for second order work was the need to establish the range of validity
of perturbation analysis. Third order calculations add nothing to
this, so they are motivated only by the possibility of higher accuracy.
But higher accuracy is guaranteed only if the perturbation series
is convergent. There is no reason to expect that for large values
of the expansion parameter the series is convergent, and for small
values the marginal increase in accuracy would not seem to justify 
the work of going to higher order computations.

\section*{Acknowledgments}

This work was supported in part by grants NSF-INT-9512894,
NSF-PHY-9423950, NSF-PHY-9507719, NATO-CRG-971092, NSF-PHY-9407194, by
funds of the University of C\'{o}rdoba, the University of Utah, the
Pennsylvania State University and its Office for Minority Faculty
Development, and the Eberly Family Research Fund at Penn State. We
acknowledge support of CONICET and CONICOR (Argentina). JP also
acknowledges support from the Alfred P. Sloan Foundation and the John
S. Guggenheim Foundation, and wishes to thank Kip Thorne for
hospitality at Caltech where this manuscript was completed. RJG is a
member of CONICET (Argentina).  We wish to thank Marco Bruni and
Alcides Garat for useful discussions and Takashi Nakamura, Misao
Sasaki and Kip Thorne for pointing out the Tomita and Tajima
references.

\end{document}